\documentclass{emulateapj} 
\usepackage{apjfonts}
\usepackage[breaklinks,colorlinks,citecolor=blue,linkcolor=magenta]{hyperref} 

\newcommand{\etal}{et~al.}
\newcommand{\PVdblt}{{\rm P}\kern 0.1em{\sc v}~$\lambda\lambda 1117, 1128$}
\newcommand{\CaIIdblt}{{\rm Ca}\kern 0.1em{\sc ii}~$\lambda\lambda 3934, 3969$}
\newcommand{\AlIIIdblt}{{\rm Al}\kern 0.1em{\sc iv}~$\lambda\lambda 1855, 1863$}
\newcommand{\CIVdblt}{{\rm C}\kern 0.1em{\sc iv}~$\lambda\lambda 1548, 1550$}
\newcommand{\MgIIdblt}{{\rm Mg}\kern 0.1em{\sc ii}~$\lambda\lambda 2796, 2803$}
\newcommand{\NVdblt}{{\rm N}\kern 0.1em{\sc v}~$\lambda\lambda 1238, 1242$}  
\newcommand{\SVIdblt}{{\rm S}\kern 0.1em{\sc vi}~$\lambda\lambda 933, 944$} 
\newcommand{\OVIdblt}{{\rm O}\kern 0.1em{\sc vi}~$\lambda\lambda 1031, 1037$} 
\newcommand{\SiIIdblt}{{\rm Si}\kern 0.1em{\sc ii}~$\lambda\lambda 1190, 1193$} 
\newcommand{\SiIVdblt}{{\rm Si}\kern 0.1em{\sc iv}~$\lambda\lambda 1393, 1402$} 
\newcommand{\PV}{\hbox{{\rm P}\kern 0.1em{\sc v}}}
\newcommand{\AlI}{\hbox{{\rm Al}\kern 0.1em{\sc i}}}
\newcommand{\AlII}{\hbox{{\rm Al}\kern 0.1em{\sc ii}}}
\newcommand{\AlIII}{{\hbox{\rm Al}\kern 0.1em{\sc iii}}}
\newcommand{\CaII}{\hbox{{\rm Ca}\kern 0.1em{\sc ii}}}
\newcommand{\CII}{\hbox{{\rm C}\kern 0.1em{\sc ii}}}
\newcommand{\CIIe}{\hbox{{\rm C$^{\ast}$}\kern 0.1em{\sc ii}}}
\newcommand{\CIII}{\hbox{{\rm C}\kern 0.1em{\sc iii}}}
\newcommand{\CIV}{\hbox{{\rm C}\kern 0.1em{\sc iv}}}
\newcommand{\CV}{\hbox{{\rm C}\kern 0.1em{\sc v}}}
\newcommand{\HI}{\hbox{{\rm H}\kern 0.1em{\sc i}}}
\newcommand{\HII}{\hbox{{\rm H}\kern 0.1em{\sc ii}}}
\newcommand{\Lya}{\hbox{{\rm Ly}\kern 0.1em$\alpha$}}
\newcommand{\Lyb}{\hbox{{\rm Ly}\kern 0.1em$\beta$}}
\newcommand{\Lyg}{\hbox{{\rm Ly}\kern 0.1em$\gamma$}}
\newcommand{\Lyd}{\hbox{{\rm Ly}\kern 0.1em$\delta$}}
\newcommand{\Lye}{\hbox{{\rm Ly}\kern 0.1em$\epsilon$}}
\newcommand{\Lyphi}{\hbox{{\rm Ly}\kern 0.1em$\phi$}}
\newcommand{\Lyfive}{\hbox{{\rm Ly}\kern 0.1em$5$}}
\newcommand{\Lysix}{\hbox{{\rm Ly}\kern 0.1em$6$}}
\newcommand{\Lyseven}{\hbox{{\rm Ly}\kern 0.1em$7$}}
\newcommand{\Lyeight}{\hbox{{\rm Ly}\kern 0.1em$8$}}
\newcommand{\Lynine}{\hbox{{\rm Ly}\kern 0.1em$9$}}
\newcommand{\Lyten}{\hbox{{\rm Ly}\kern 0.1em$10$}}
\newcommand{\Lyeleven}{\hbox{{\rm Ly}\kern 0.1em$11$}}
\newcommand{\HeI}{\hbox{{\rm He}\kern 0.1em{\sc i}}}
\newcommand{\HeII}{\hbox{{\rm He}\kern 0.1em{\sc ii}}}
\newcommand{\FeI}{\hbox{{\rm Fe}\kern 0.1em{\sc i}}}
\newcommand{\FeII}{\hbox{{\rm Fe}\kern 0.1em{\sc ii}}}
\newcommand{\FeIII}{\hbox{{\rm Fe}\kern 0.1em{\sc iii}}}
\newcommand{\MnII}{\hbox{{\rm Mn}\kern 0.1em{\sc ii}}}
\newcommand{\MgI}{\hbox{{\rm Mg}\kern 0.1em{\sc i}}}
\newcommand{\MgIb}{\hbox{{\rm Mg}\kern 0.1em{\sc i}}\kern 0.05em{\rm b}}
\newcommand{\MgII}{\hbox{{\rm Mg}\kern 0.1em{\sc ii}}}
\newcommand{\MgIII}{\hbox{{\rm Mg}\kern 0.1em{\sc iii}}}
\newcommand{\NI}{\hbox{{\rm N}\kern 0.1em{\sc i}}}
\newcommand{\NII}{\hbox{{\rm N}\kern 0.1em{\sc ii}}}
\newcommand{\NIII}{\hbox{{\rm N}\kern 0.1em{\sc iii}}}
\newcommand{\NV}{\hbox{{\rm N}\kern 0.1em{\sc v}}}
\newcommand{\OVI}{\hbox{{\rm O}\kern 0.1em{\sc vi}}}
\newcommand{\OI}{\hbox{{\rm O}\kern 0.1em{\sc i}}}
\newcommand{\OII}{\hbox{[{\rm O}\kern 0.1em{\sc ii}]}}
\newcommand{\OIII}{\hbox{[{\rm O}\kern 0.1em{\sc iii}]}}
\newcommand{\OIV}{\hbox{{\rm O}\kern 0.1em{\sc iv}]}}
\newcommand{\SI}{{\rm S}\kern 0.1em{\sc i}}
\newcommand{\SIV}{{\rm S}\kern 0.1em{\sc iv}}
\newcommand{\SVI}{{\rm S}\kern 0.1em{\sc vi}}
\newcommand{\SiI}{\hbox{{\rm Si}\kern 0.1em{\sc i}}}
\newcommand{\SiII}{\hbox{{\rm Si}\kern 0.1em{\sc ii}}}
\newcommand{\SiIII}{\hbox{{\rm Si}\kern 0.1em{\sc iii}}}
\newcommand{\SiIV}{\hbox{{\rm Si}\kern 0.1em{\sc iv}}}
\newcommand{\SII}{\hbox{{\rm S}\kern 0.1em{\sc ii}}}
\newcommand{\SIII}{\hbox{{\rm S}\kern 0.1em{\sc iii}}}
\newcommand{\NaI}{\hbox{{\rm Na}\kern 0.1em{\sc i}}}
\newcommand{\NaID}{\hbox{{\rm Na}\kern 0.1em{\sc i}}\kern 0.05em{\rm D}}
\newcommand{\TiII}{\hbox{{\rm Ti}\kern 0.1em{\sc ii}}}
\newcommand{\kms}{\hbox{~km~s$^{-1}$}}

\slugcomment{} 
\shorttitle{\sc Azimuthal dependence of {\OVI} absorption}
\shortauthors{\sc Kacprzak et~al.}

\begin{document}


\title{The Azimuthal dependence of outflows and accretion detected using {\OVI} absorption}


\author{\sc Glenn G. Kacprzak\altaffilmark{1}, Sowgat
  Muzahid\altaffilmark{2}, Christopher W. Churchill\altaffilmark{3},
  Nikole M. Nielsen\altaffilmark{1}, Jane C. Charlton\altaffilmark{2}}

\altaffiltext{1}{Swinburne University of Technology, Victoria 3122, Australia {\tt gkacprzak@astro.swin.edu.au}}
\altaffiltext{2}{The Pennsylvania State University, State College, PA 16801, USA}
\altaffiltext{3}{New Mexico State University, Las Cruces, NM 88003, USA}

\begin{abstract}
 We report a bimodality in the azimuthal angle ($\Phi$) distribution of
gas around galaxies traced by {\OVI} absorption. We present the mean
$\Phi$ probability distribution function of 29~{\it HST}-imaged {\OVI}
absorbing (EW>0.1~{\AA}) and 24~non-absorbing (EW<0.1~{\AA}) isolated
galaxies (0.08$<$$z$$<$0.67) within $\sim200$~kpc of background
quasars. We show that EW is anti-correlated with impact parameter and
{\OVI} covering fraction decreases from 80\% within 50~kpc to 33\% at
200~kpc.  The presence of {\OVI}~absorption is azimuthally dependent
and occurs between $\pm$10--20$^{\circ}$ of the galaxy projected major
axis and within $\pm$30$^{\circ}$ of the projected minor axis.  We
find higher EWs along the projected minor axis with weaker EWs along
the project major axis. Highly inclined galaxies have the lowest
covering fractions due to minimized outflow/inflow cross-section
geometry. Absorbing galaxies also have bluer colors while
non-absorbers have redder colors, suggesting that star-formation is a
key driver in the {\OVI}~detection rate. {\OVI}~surrounding blue
galaxies exists primarily along the projected minor axis with wide
opening angles while {\OVI}~surrounding red galaxies exists primarily
along the projected major axis with smaller opening angles, which may
explain why absorption around red galaxies is less frequently
detected. Our results are consistent with CGM originating from major
axis-fed inflows/recycled gas and from minor axis-driven
outflows. Non-detected {\OVI}~occurs between $\Phi$=20--60$^{\circ}$,
suggesting that {\OVI}~is not mixed throughout the CGM and remains
confined within the outflows and the disk-plane. We find low {\OVI}
covering fractions within $\pm10$$^{\circ}$ of the projected major
axis, suggesting that cool dense gas resides in a narrow planer
geometry surrounded by diffuse {\OVI}~gas. 
\end{abstract}




%

\section{Introduction}
\label{sec:intro}

The circumgalactic medium (CGM) supplies the gas reservoirs required
for galaxy star formation and is replenished via accretion/recycling
and outflowing gas. With the likelihood that the CGM can make up about
50\% of the baryonic mass bound to galaxies \citep{tumlinson11} and
could consist of about 50\% of the baryons unaccounted for around
galaxies \citep{werk14}, the CGM's importance in dictating galaxy
evolution cannot be overstated. Determining how the CGM interacts with
galaxies is critical to understanding how galaxies evolve.

Gas accretion from the intergalactic medium, along with previously
ejected recycled gas, and large-scale galactic outflows are principal
components of theoretical galaxy formation and evolution models
\citep{springel03,keres05,dekel09,oppenheimer10,
  dave11a,dave11b,stewart11b,dave12,ford14}. Many studies have shown
evidence for the existence of both multi-phase cold
accretion/recycling
\citep[e.g.,][]{steidel02,zonak04,chen05,tripp05,cooksey08,chen10a,chen10b,kacprzak10,kacprzak11b,ribaudo11,thom11,kacprzak12b,martin12,noterdaeme12,rubin12,bouche13,crighton13,krogager13,peroux13}
and large-scale galactic outflows
\citep[e.g.,][]{bouche06,tremonti07,zibetti07,martin09,weiner09,chelouche10,nestor11,noterdaeme10,bordoloi11,coil11,rubin10,menard12,martin12,noterdaeme12,krogager13,peroux13,kacprzak14,lan14,rubin14,crighton15,muzahid15}. 

Both models and observations indicated that gas accretion should occur
along filaments co-planar to the galaxy disk, whereas outflows are
expected to extend along the galaxy projected minor axis.  Evidence
for the geometric preference of inflowing and outflowing gas has
already been observed for cool gas traced by {\MgII} absorption
\citep{kacprzak11b,bouche12,bordoloi11,kacprzak12a,lan14,bordoloi14}. It
was originally reported that the {\MgII} equivalent width is dependent
on galaxy inclination \citep{kacprzak11b} with the absorption
kinematics consistent with being coupled to the galaxy angular
momentum suggesting co-planar geometry
\citep{steidel02,kacprzak10,diamond-stanic15}.  \citet{kacprzak12a}
reported a bimodality in the azimuthal angle distribution of gas
around galaxies, where cool ($T\sim$10$^4$~K) dense CGM gas prefers to
exist along the projected galaxy major and minor axes \citep[also
see][]{bouche12,bordoloi11,lan14} with the gas covering fraction being
enhanced by as much as 20\%--30\% along these axes.  They found that
blue star-forming galaxies drive the bimodality while red passive
galaxies contain gas along their projected major axis. Outflows likely
contain more metal enriched gas and higher velocity width absorption
profiles since higher equivalent width absorption tends to reside
along the projected minor axes of galaxies
\citep{bordoloi11,kacprzak12a,lan14}. \citet{nielsen15} has shown that
{\MgII} absorption profiles with the largest velocity dispersion are
associated with blue, face-on galaxies probed along the projected
minor axis while the cloud column densities are largest for edge-on
galaxies and blue galaxies.  \citet{bouche12} have also shown that
absorption detected along the projected minor axes is kinematically
consistent with bi-conical outflows.  These combined results are
consistent with galaxy evolution scenarios where star-forming galaxies
accrete new cool/warm co-planer gas within an half-opening angle of
about 20$^{\circ}$, forming new stars and producing metal-enriched
galactic scale outflows with half opening angles of 50$^{\circ}$,
while red galaxies exist passively due to reduced gas reservoirs
\citep{kacprzak12a}. The conclusions are drawn from observations
conducted using {\MgII} absorption, however both outflowing and
infalling gas are expected to be mulit-phased and should be explored
using other gas-phase tracers such as {\CIV}, {\OVI}, etc.

Another standard tracer of the diffuse CGM gas is the {\OVIdblt}
doublet. We know that there is a significant fraction of {\OVI}
contained in the CGM
\citep{stocke06,tumlinson11,stocke13,peeples14,werk14} that is
typically gravitationally bound within the host galaxy's gravitational
potential \citep{tumlinson11,stocke13,mathes14}. Compared to {\MgII},
the {\OVI} doublet can be more difficult to interpret since it can be
commonly detected as photo-ionized or collisionally ionized gas. This
implies that {\OVI} can trace warm/hot coronal regions surrounding
galaxies, which may dictate the formation and destruction of the
cool/warm CGM \citep{mo96,maller04,dekel06} or trace other multi-phase
gas structures. Although {\OVI} absorption has been extensively
studied in and around galaxies \citep{savage03,sembach04,stocke06,
  danforth08,
  tripp08,wakker09,prochaska11,tumlinson11,johnson13,stocke13,mathes14,johnson15},
the geometrical distribution of {\OVI} absorption around galaxies,
which can yield improved understanding of the origins of such gas, has
not been thoroughly observationally studied.

\citet{mathes14} attempted to address the azimuthal dependence of
{\OVI} using 14 galaxies and found a mostly spatially uniform
distribution of absorbing gas out to 300~kpc.  There was small hint of
a bimodality but was based on only five galaxies that were probed
within one viral radius.

We aim to further explore the multi-phase inflow and outflow azimuthal
distribution using {\OVI} absorption for a large sample of
spectroscopically confirmed galaxies. In Section~\ref{sec:data} we
present our sample and data reduction. In Section~\ref{sec:results} we
present the equivalent width (EW) and covering fraction dependence on
impact paramater, D, and we compute the mean azimuthal angle
probability distribution function for absorbing (EW>0.1~{\AA}) and
non-absorbing (EW<0.1~{\AA}) galaxies along with the azimuthal angle
covering fraction. We further compare the colors for absorbing and
non-absorbing galaxies and show the azimuthal angle probability
distribution function depends on galaxy color.  In
Section~\ref{sec:discussion}, we discuss what can be inferred from the
results and concluding remarks are offered in
Section~\ref{sec:conclusion}. Throughout we adopt an H$_{\rm
  0}=70$~\kms Mpc$^{-1}$, $\Omega_{\rm M}=0.3$, $\Omega_{\Lambda}=0.7$
cosmology.

\begin{figure}
\begin{center}
\includegraphics[angle=0,scale=0.56]{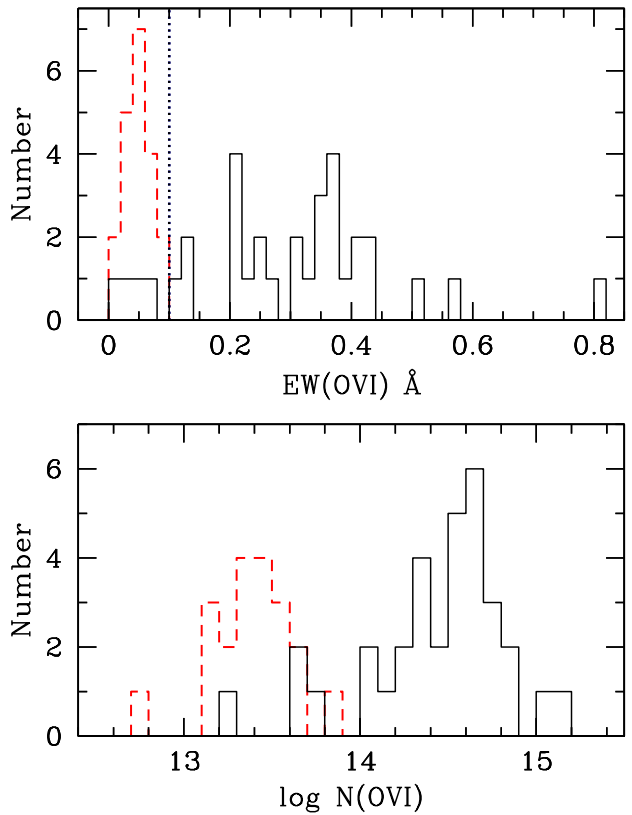}
\caption[angle=0]{(Top) Distribution of {\OVI} rest-frame equivalent
  widths (solid black) and 3$\sigma$ limits (dashed red). We have
  adopted a {\OVI} detection threshold of $\geq$0.1~{\AA} given the
  distribution of equivalent widths. Detections below this threshold
  are considered as non-absorbers such that our sample is treated with
  the same level of sensitivity. (Bottom) The {\OVI} column density
  distribution is shown for the same sample. Our equivalent
  width bifurcation at 0.1~{\AA} translates to a column density
  bifurcation log $N$({\OVI})=14.0.}
\label{fig:limits}
\end{center}
\end{figure}

\begin{figure*}
\begin{center}
\includegraphics[scale=0.59]{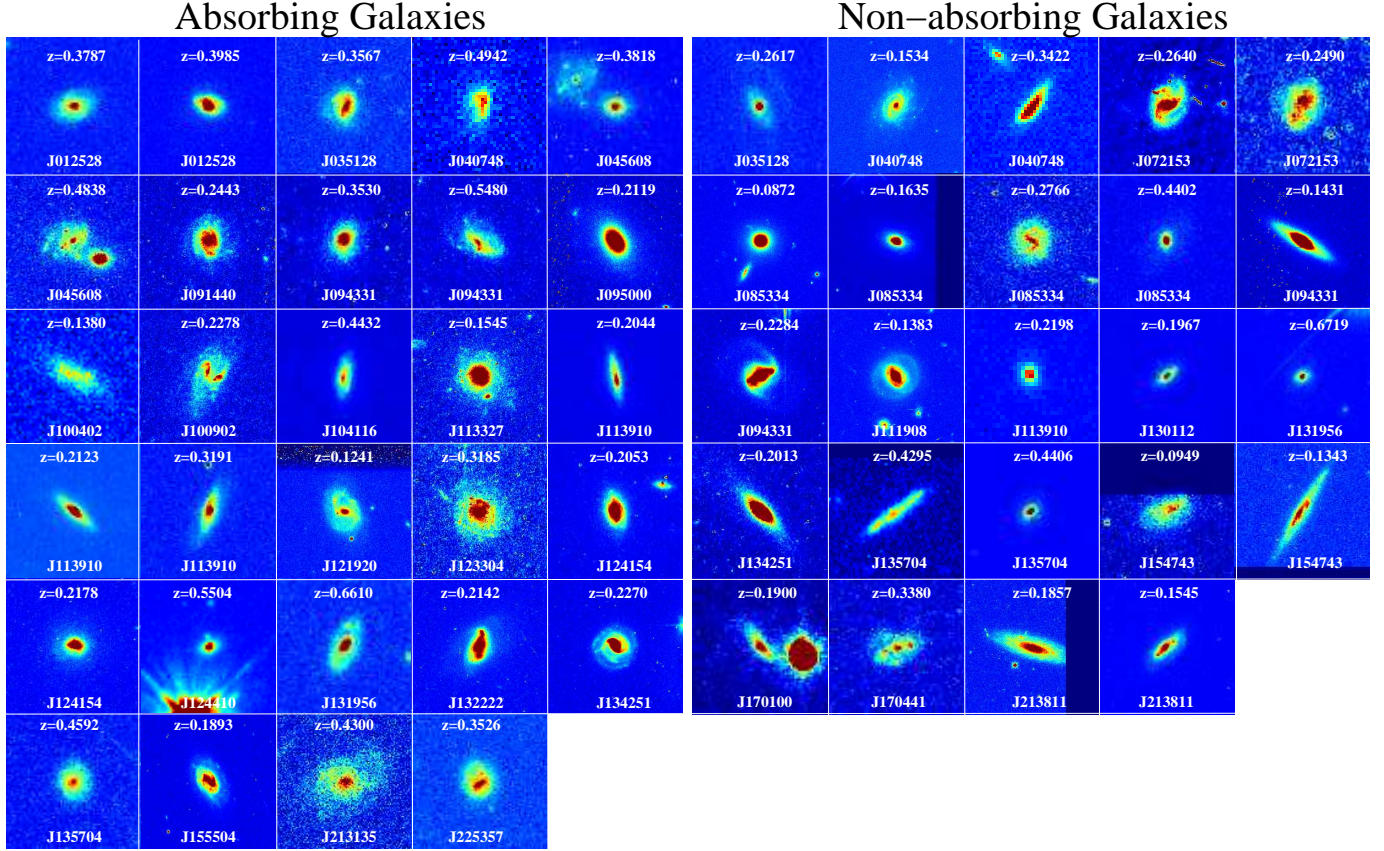}
\caption[angle=0]{ (Left) {\it HST} images of 29 absorbing galaxies
  selected by {\OVI} absorption (EW>0.1{\AA}). The images are 10 times
  larger than the $1.5\sigma$ isophotal area. The image orientation is
  set by the parent {\it HST} image and is arbitrary. --- (Right) same
  as left except for 24 non-absorbing galaxies selected with
  EW<0.1{\AA}. }
\label{fig:absimage}
\end{center}
\end{figure*}

\section{GALAXY SAMPLE AND DATA ANALYSIS}
\label{sec:data}

We have constructed a sample of 53 candidate {\OVI} absorbing galaxies
with spectroscopically confirmed redshifts ranging between
0.08$<$$z$$<$0.67 ($\left< z \right> = 0.29$) within $\sim200$~kpc
(21$<$$z$$<$203~kpc) of bright background quasars. These galaxies are
selected to be isolated such that there are no neighbors within
100~kpc and have velocity separations less than 500~{\kms}. All these
galaxy--absorber pairs were identified as part of our ``Multiphase
Galaxy Halos'' Survey (from PID 13398 plus from the literature).  The
quasars and quasar fields are selected to have {\it HST} imaging and
medium resolution {\it HST} ultraviolet spectra. We discuss the data
and analysis below.

\subsection{Quasar Spectroscopy}

We have compiled 32 quasars from our survey that have medium
resolution ($R\sim$20,000) spectra that cover the {\OVIdblt} doublet
for the targeted galaxies. Details of the {\it HST}/COS and STIS
observations are contained in Table~\ref{tab:HST}.

The data were reduced using the {\sc CALCOS} pipeline software. The
pipeline reduced data (`x1d' files) were flux calibrated. In order to
increase the spectral signal-to-noise ratio ($S/N$), individual
grating integrations were aligned and co-added using the {\sc IDL}
code `coadd\_x1d' developed by
\citet{danforth10}\footnote{http://casa.colorado.edu/danforth/science/cos/costools.html}. Since
the archival spectra come from different observing programs, our
sample shows a range in $S/N$ from 5--25 per resolution element. As
the COS FUV spectra are significantly over-sampled, i.e., six raw
pixels per resolution element, we binned the data by three
pixels. Binning of data improves $S/N$ per pixel by a factor of $\sqrt
3$. All the measurements and analysis were performed on the binned
spectra. Continuum normalizations were done by fitting the line-free
regions with smooth low-order polynomials.

The line spread function (LSF) of the COS spectrograph is not a
Gaussian. For our Voigt profile analysis, we adopt the non-Gaussian
LSF given by \citet{kriss11}. The LSF was obtained by interpolating
the LSF tables at the observed central wavelength for each absorption
line and was convolved with the model Voigt profile while fitting
absorption lines using {\sc VPFIT}\footnote{http://www.ast.cam.ac.uk/
  rfc/vpfit.html} software. Whenever possible, for each identified
system, both the members of {\OVIdblt} doublet were fitted
simultaneously in order to get best fitting component column
densities. In cases for which one of the doublets is affected by a
blend, we used the unblended transition to constrain the fit
parameters. We then generate synthetic profile(s) for the blended
component(s) for a consistency check. In all cases, we have used the
minimum number of components required to get a satisfactory fit with
reduced $\chi^2 \sim 1$. The model profiles of $\lambda$~1031 were used
to compute the {\OVI} equivalent width (EW). Errors and 3$\sigma$
limits on the EWs were computed using the error spectrum. The EWs and
column densities are listed in Table~\ref{tab:morph}.

In Figure~\ref{fig:limits}, we show the distribution of
\OVI$\lambda$1031 rest-frame equivalent widths and 3$\sigma$
equivalent width upper-limits along with the column density
distribution.  There is a clear transition between absorbing and
non-absorbing galaxies at 0.1~{\AA} with only four absorbers detected
below this threshold. In order to treat our data/sample with uniform
sensitivity limits, we consider only stronger systems with
EW>0.1~{\AA} as absorbers and all galaxies below this threshold as
non-absorbers. The 0.1~{\AA} bifurcation translates to a column
density cut of log~$N$(\OVI)$\simeq$14. Our final sample consists of 29
absorbing and 24 non-absorbing galaxies.

\begin{deluxetable*}{llllclcllrr}
\tabletypesize{\scriptsize}
\tablecaption{ {\it HST\/} UV Spectra and Optical Imaging Observations\label{tab:HST}}
\tablecolumns{11}
\tablewidth{0pt} 
\tablehead{
\colhead{QSO}&
\colhead{$z_{\rm QSO}$} &
\colhead{RA$_{\rm QSO}$} &
\colhead{DEC$_{\rm QSO}$}&
\colhead{UV} &
\colhead{UV} &
\colhead{PID} &
\colhead{Imaging} &
\colhead{Filter}&
\colhead{Exposure} &
\colhead{PID }\\
\colhead{ }&
\colhead{ } &
\colhead{(J2000)} &
\colhead{(J2000)}&
\colhead{Camera } &
\colhead{Grating} &
\colhead{ } &
\colhead{Camera} &
\colhead{ }&
\colhead{(sec) } & 
\colhead{ }
}
\startdata
J012528.84$-$000555.9  & 1.075  &  01:25:28.84 &  $-$00:05:55.93 & COS  &  G160M         &    13398      & WFPC2   & F702W  &  700   & 6619  \\ 
J022815.17$-$405714.3  & 0.493  &  02:28:15.17 &  $-$40:57:14.29 & COS  &  G130M, G160M  &    11541      & ACS        & F814W  &  1200  & 13024 \\ 
J035128.54$-$142908.7  & 0.616  &  03:51:28.54 &  $-$14:29:08.71 & COS  &  G130M, G160M  &    13398      & WFPC2   & F702W  &  800   & 5949  \\ 
J040748.44$-$121136.8  & 0.573  &  04:07:48.43 &  $-$12:11:36.66 & COS  &  G130M, G160M  &    11541      & WFPC2   & F702W  &  800   & 5949  \\ 
J045608.92$-$215909.4  & 0.534  &  04:56:08.92 &  $-$21:59:09.40 & COS  &  G160M         &  12466,12252  & WFPC2   & F702W  &  600   & 5098  \\ 
J072153.44$+$712036.3  & 0.300  &  07:21:53.45 &  $+$71:20:36.36 & COS  &  G130M, G160M  &    12025      & ACS        & F814W  &  1200  & 13024 \\ 
J085334.25$+$434902.3  & 0.515  &  08:53:34.25 &  $+$43:49:02.33 & COS  &  G130M, G160M  &    13398      & WFPC2   & F702W  &  800   & 5949  \\ 
J091440.39$+$282330.6  & 0.735  &  09:14:40.39 &  $+$28:23:30.62 & COS  &  G130M, G160M  &    11598      & ACS        & F814W  &  1200  & 13024 \\ 
J094331.61$+$053131.4  & 0.564  &  09:43:31.62 &  $+$05:31:31.49 & COS  &  G130M, G160M  &    11598      & ACS        & F814W  &  1200  & 13024 \\ 
J095000.73$+$483129.3  & 0.590  &  09:50:00.74 &  $+$48:31:29.38 & COS  &  G130M, G160M  &    11598      & ACS        & F814W  &  1200  & 13024 \\ 
J100402.61$+$285535.4  & 0.327  &  10:04:02.61 &  $+$28:55:35.39 & COS  &  G130M, G160M  &    12038      & WFPC2   & F702W  &  800   & 5949  \\ 
J100902.07$+$071343.9  & 0.457  &  10:09:02.07 &  $+$07:13:43.87 & COS  &  G130M, G160M  &    11598      & WFC3    & F625W  &  2256  & 11598 \\ 
J104116.16$+$061016.9  & 1.270  &  10:41:17.16 &  $+$06:10:16.92 & COS  &  G160M         &    12252      & WFPC2   & F702W  &  1300  & 5984  \\ 
J111908.67$+$211918.0  & 0.177  &  11:19:08.68 &  $+$21:19:18.01 & COS  &  G130M, G160M  &    12038      & WFPC2   & F606W  &  2200  & 5849  \\ 
J113327.78$+$032719.1  & 0.525  &  11:33:27.79 &  $+$03:27:19.17 & COS  &  G130M, G160M  &    11598      & ACS        & F814W  &  1200  & 13024 \\ 
J113910.79$-$135043.6  & 0.557  &  11:39:10.70 &  $-$13:50:43.64 & COS  &  G130M         &    12275      & ACS     & F814W  &  520   & 9682  \\ 
J121920.93$+$063838.5  & 0.331  &  12:19:20.93 &  $+$06:38:38.52 & COS  &  G130M, G160M  &    12025      & WFPC2   & F702W  &  600   & 5143  \\ 
J123304.05$-$003134.1  & 0.471  &  12:33:04.05 &  $-$00:31:34.20 & COS  &  G130M, G160M  &    11598      & ACS     & F814W  &  1200  & 13024 \\ 
J124154.02$+$572107.3  & 0.584  &  12:41:54.02 &  $+$57:21:07.38 & COS  &  G130M, G160M  &    11598      & ACS        & F814W  &  1200  & 13024 \\ 
J124410.82$+$172104.5  & 1.273  &  12:44:10.82 &  $+$17:21:04.52 & COS  &  G160M         &    12466      & WFPC2   & F702W  &  1300  & 6557  \\ 
J130112.93$+$590206.7  & 0.478  &  13:01:12.93 &  $+$59:02:06.75 & COS  &  G130M, G160M  &    11541      & WFPC2   & F702W  &  700   & 6619  \\ 
J131956.23$+$272808.2  & 1.015  &  13:19:56.23 &  $+$27:28:08.22 & COS  &  G160M         &    11667      & WFPC2   & F702W  &  1300  & 5984  \\ 
J132222.46$+$464546.1  & 0.375  &  13:22:22.68 &  $+$46:45:35.22 & COS  &  G130M, G160M  &    11598      & ACS        & F814W  &  1200  & 13024 \\ 
J134251.60$-$005345.3  & 0.327  &  13:42:51.61 &  $-$00:53:45.31 & COS  &  G130M, G160M  &    11598      & ACS        & F814W  &  1200  & 13024 \\ 
J135704.43$+$191907.3  & 0.720  &  13:57:04.43 &  $+$19:19:07.37 & COS  &  G160M         &    13398      & WFPC2   & F702W  &  800   & 5949  \\ 
J154743.53$+$205216.6  & 0.264  &  15:47:43.53 &  $+$20:52:16.61 & COS  &  G130M, G160M  &    13398      & WFPC2   & F702W  &  1100  & 5099  \\ 
J155504.39$+$362847.9  & 0.714  &  15:55:04.40 &  $+$36:28:48.04 & COS  &  G130M, G160M  &    11598      & ACS        & F814W  &  1200  & 13024 \\ 
J170100.60$+$641209.3  & 2.741  &  17:01:00.62 &  $+$64:12:09.12 & COS  &  G130M         &    13491      & ACS        & F814W  &  12520 & 10581 \\ 
J170441.37$+$604430.5  & 0.372  &  17:04:41.38 &  $+$60:44:30.50 & STIS &  E140M         &    8015       & WFPC2   & F702W  &  600   & 5949  \\ 
J213135.26$-$120704.8  & 0.501  &  21:31:35.26 &  $-$12:07:04.79 & COS  &  G160M         &    13398      & WFPC2   & F702W  &  600   & 5143  \\ 
J213811.60$-$141838.0  & 1.900  &  21:38:11.60 &  $-$14:18:38.0  & COS  &  G130M, G160M  &    13398      & WFPC2   & F702W  &  1400  & 5343  \\ 
J225357.74$+$160853.6  & 0.859  &  22:53:57.74 &  $+$16:08:53.56 & COS  &  G130M, G160M  &    13398      & WFPC2   & F702W  &  700   & 6619  
\enddata 
\end{deluxetable*} 

\subsection{HST Imaging of Quasar Fields}

We further required that all 32 quasar fields were imaged with
high-resolution spaced-based cameras such that we were able to
adequately model the morphological parameters and orientations of all
53 galaxies.  Details of the {\it HST} imaging observations using ACS,
WFC3, and WFPC2 for a range of filters are listed in
Table~\ref{tab:HST}.

The WFPC--2/{\it HST\/} images were reduced using the WFPC--2
Associations Science Products Pipeline (WASPP)
\citep[see][]{kacprzak11b}. The ACS and WFC3 data reduction was carried
out using the DrizzlePac software \citep{gonzaga12}. If enough frames
were present, cosmic rays were removed during the multidrizzle process
otherwise we used lacosmic \citep{vandokkum01}.

Galaxy photometry was performed using the Source Extractor package
\citep[SExtractor;][]{bertin96} with a detection criterion of
1.5~$\sigma$ above background.  The $m_{HST}$ magnitudes in each
filter were measured using the WFPC--2 Vega zeropoints
\citep{whitmore95}, which were then converted to AB magnitudes
\citep[see][]{nielsen13a}, while the ACS, and WFC3 zero points are
based upon the AB system. The $m_{HST}$, and its corresponding filter,
for each galaxy is listed in Table~\ref{tab:morph}.

\subsection{GIM2D Galaxy Models}

Galaxy morphological parameters and orientations were quantified by
fitting a two-component (bulge+disk) model using GIM2D \citep[Galaxy
IMage 2D;][]{simard02}.  We fit the surface brightness of the disk
component with an exponential profile and we fit the bulge component
with a S{\'e}rsic profile \citep{sersic68} where the S{\'e}rsic index
may vary between $0.2\leq n\leq 4.0$. This technique has been
successfully applied in previous works
\citep{kacprzak07,kacprzak11b,kacprzak12a}. The GIM2D outputs were
manually inspected to see if models were realistic representations of
the observed galaxies.

To model the galaxies, GIM2D extracts ``portrait size'' images from
parent {\it HST} images with an area 10 times larger than the
$1.5\sigma$ galaxy isophotal area such that an accurate background can
be computed. Figure~\ref{fig:absimage} shows the portraits of the
absorbing and non-absorbing galaxies. All galaxies appear to have
similar sizes since they are shown for an area 10 times larger than
the $1.5\sigma$ galaxy isophotal area. The orientations of the images
are that of the parent {\it HST} and are arbitrary. Note there is a
wide range of galaxy morphological types with the dominant population
being disk galaxies.

During the GIM2D process, the models are convolved with the point
spread function (PSF). For WFPC-2, we modeled the PSF at the
appropriate locations on the WFPC-2 chip using Tiny Tim
\citep{krist04} as performed by \citet{kacprzak11b}. The PSFs for WFC3
and ACS depend both on time and position on the chip and, in addition,
the images also contain significant geometrical distortions.  So for
WFC3 and ACS, we used Tiny Tim to create the PSFs and place them into
blank frames every 500 pixels, with the same size and header
parameters as those of the real flat-fielded individual exposures
which were then reduced following the same procedure as for the
data. The PSF that was closest to the galaxy of interest was used for
the GIM2D modeling.

The galaxy properties are listed in Table~\ref{tab:morph}. We adopt
the convention of the azimuthal angle $\Phi=0^{\circ}$ to be along the
galaxy major axis and $\Phi=90^{\circ}$ to be along the galaxy minor
axis. We have included in our analysis the full range of galaxy
inclinations present in our sample.  The sample contains only three
galaxies with $i<20$ degrees. We find that the inclusion or exclusion
of these three galaxies does not change our main results.  We also
note that, in the case of {\MgII} absorption, the geometric
distribution of the low-ionization CGM is only weakly dependent upon
galaxy inclination \citep{bordoloi14, kacprzak11b}, but is strongly
dependent upon azimuthal angle
\citep{bouche12,bordoloi11,kacprzak12a,lan14}. The combined $\Phi$
distribution of absorbers and non-absorbers is shown to be consistent
with a uniform random distribution using a KS test with P(KS)=0.635

The impact parameters, $D$, are computed using the galaxy and quasar
isophotal centroids determined by SExtractor. The uncertainty in $D$
is computed from the pixel offset of the galaxy SExtractor isophotal
center and the center of the GIM2D galaxy model, which is typically
$0.25$ pixels. An additional $\sim$0.05 pixel uncertainty is included
for centroiding error of the quasar, which is based upon centroiding
errors of unresolved sources in the our images.

\begin{figure}
\begin{center}
\includegraphics[angle=0,scale=0.61]{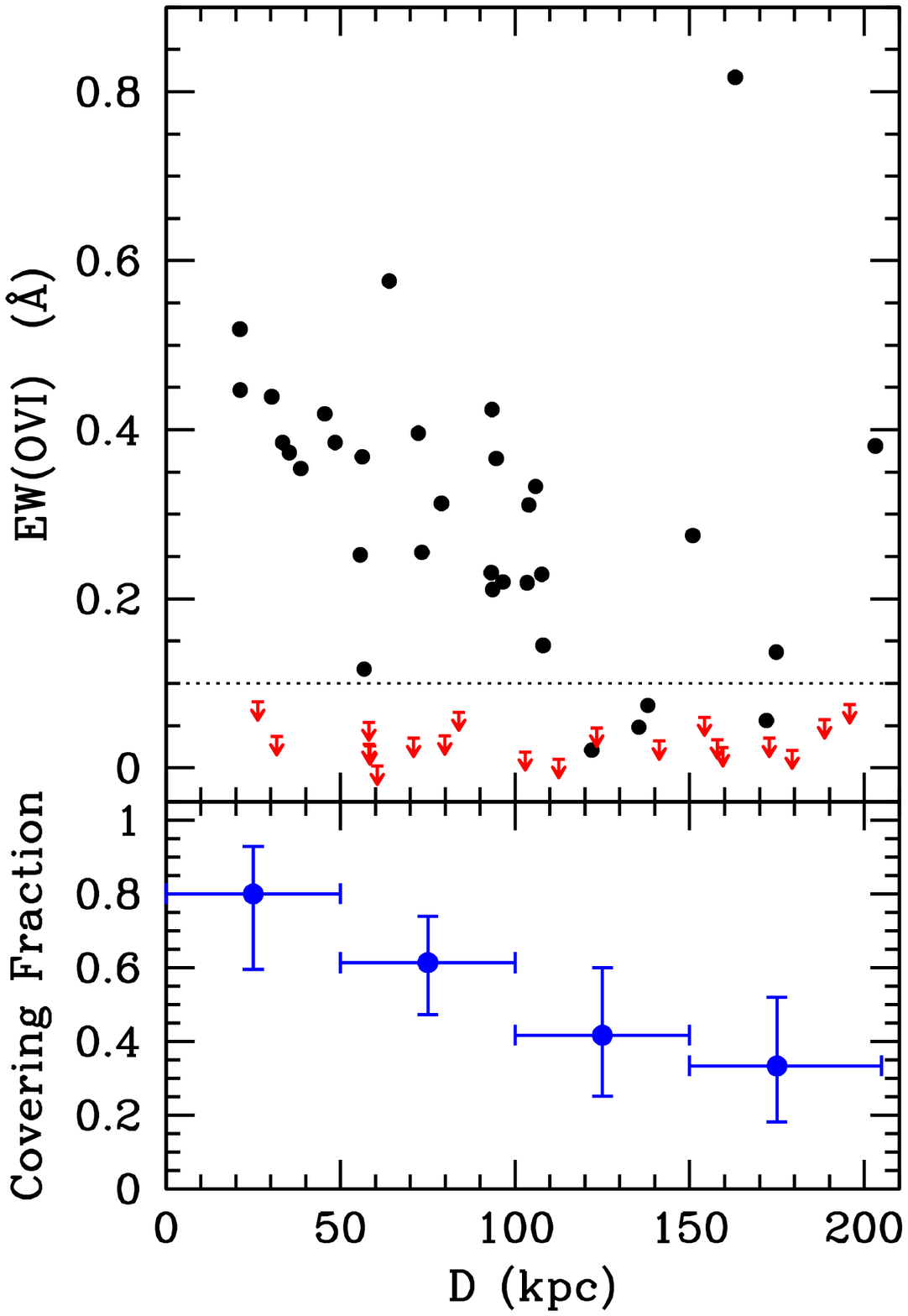}
\caption[angle=0]{(Top) Rest-frame {\OVI}$\lambda$1031 equivalent
  width versus impact parameter with 3$\sigma$ upper limits shown as
  red arrows. The horizontal dashed line shows bifurcation value of
  0.1~{\AA} separating absorbing and non-absorbing galaxies. There are
  four detections below this limit, however they are treated as
  non-detections for our analysis but shown for completeness. ---
  (Bottom) The covering fraction as a function of impact parameter for
  an equivalent sensitivity of 0.1~{\AA}. The horizontal error bars
  indicate the full range in impact parameter within each bin. The
  covering fraction 1$\sigma$ errors are derived from binomial
  statistics \citep{gehrels86}.}
\label{fig:D}
\end{center}
\end{figure}

\section{Results}\label{sec:results}

\begin{deluxetable*}{lllcccccrrrrr}
\tabletypesize{\scriptsize}
\tablecaption{Absorption and host galaxy properties\label{tab:morph}}
\tablecolumns{13}
\tablewidth{0pt} 
\tablehead{
\colhead{QS0}&
\colhead{RA$_{\rm gal}$} &
\colhead{DEC$_{\rm gal}$} &
\colhead{$z_{\rm gal}$}&
\colhead{$z_{\rm gal}$} &
\colhead{HST} &
\colhead{$m_{HST}$} &
\colhead{($B-K$)$^{\dagger}$} &
\colhead{$D$} &
\colhead{$\Phi$} &
\colhead{$i$} &
\colhead{EW} &
\colhead{log $N$({\OVI})}\\
\colhead{field }&
\colhead{(J2000)} &
\colhead{(J2000)} &
\colhead{}&
\colhead{ref\tablenotemark{ a}} &
\colhead{Filter} &
\colhead{(AB)} &
\colhead{($g-r$)$^{\spadesuit}$} &
\colhead{(kpc)} &
\colhead{(degree)} & 
\colhead{(degree)} & 
\colhead{(\AA)} & 
\colhead{ }
}
\startdata
\cutinhead{Non-Absorbing Galaxies with EW({\OVI})<0.1~\AA}
J035128 & 03:51:28.933 & $-$14:29:54.31 & 0.2617 & 1  &  F702W  & 21.0 & 2.3$^{\dagger}$    & $188.6\pm0.3$ & $64.9_{-15.8}^{+21.1}$ &$83.0_{-3.0}^{+2.0}$  & $<0.057        $ & $<13.66        $\\[+0.3ex]             
J040748 & 04:07:48.481 & $-$12:12:11.13 & 0.3422 & 2  &  F702W  & 21.5 &$\cdots$           & $172.0\pm0.1$ & $48.1_{-0.9}^{+1.0}$   &$85.0_{-0.4}^{+0.1}$  & $ 0.056\pm0.002$ & $ 13.68\pm0.05 $\\[+0.3ex]             
J040748 & 04:07:43.930 & $-$12:12:08.49 & 0.1534 & 2  &  F702W  & 18.5 &$\cdots$           & $195.9\pm0.1$ & $26.3_{-1.0}^{+0.9}$   &$49.5_{-0.7}^{+0.5}$  & $<0.075        $ & $<12.78        $\\[+0.3ex]             
J072153 & 07:21:51.403 & $+$71:20:10.80 & 0.2640 & 3  &  F814W  & 19.0 &$\cdots$           & $112.5\pm0.1$ & $1.5_{-0.1}^{+0.3}$    &$49.9_{-0.1}^{+0.2}$  & $<0.010        $ & $<13.20        $\\[+0.3ex]             
J072153 & 07:21:54.962 & $+$71:20:11.20 & 0.2490 & 3  &  F814W  & 22.1 &$\cdots$           & $102.9\pm0.1$ & $38.8_{-4.9}^{+6.9}$   &$79.3_{-2.3}^{+5.3}$  & $<0.019        $ & $<13.17        $\\[+0.3ex]             
J085334 & 08:53:35.160 & $+$43:48:59.81 & 0.4402 & 1  &  F702W  & 20.6 & 1.8$^{\dagger}$    & $58.1\pm0.4$  & $23.0_{-7.6}^{+6.5}$   &$73.3_{-3.0}^{+3.8}$  & $<0.028        $ & $<13.35        $\\[+0.3ex]             
J085334 & 08:53:33.384 & $+$43:49:03.97 & 0.1635 & 1  &  F702W  & 18.8 & 1.8$^{\dagger}$    & $26.2\pm0.1$  & $56.0_{-0.8}^{+0.8}$   &$70.1_{-0.8}^{+1.4}$  & $<0.078        $ & $<13.87        $\\[+0.3ex]     
J085334 & 08:53:36.881 & $+$43:49:33.32 & 0.2766 & 1  &  F702W  & 20.9 &$\cdots$           & $179.4\pm0.2$ & $36.7_{-15.3}^{+14.9}$ &$32.8_{-6.7}^{+5.7}$  & $<0.021        $ & $<13.53        $\\[+0.3ex]             
J085334 & 08:53:34.481 & $+$43:49:37.51 & 0.0872 & 1  &  F702W  & 17.5 & 2.1$^{\dagger}$    & $58.4\pm0.1$  & $50.1_{-1.3}^{+1.5}$   &$30.9_{-0.7}^{+0.5}$  & $<0.026        $ & $<13.63        $\\[+0.3ex]             
J094331 & 09:43:29.210 & $+$05:30:41.75 & 0.1431 & 4  &  F814W  & 17.5 & 2.82$^{\spadesuit}$ & $154.2\pm0.1$ & $77.7_{-0.1}^{+0.1}$  &$75.5_{-0.1}^{+0.1}$  & $<0.060        $ & $<13.68        $\\[+0.3ex]             
J094331 & 09:43:33.789 & $+$05:31:22.26 & 0.2284 & 4  &  F814W  & 18.6 & 2.24$^{\spadesuit}$ & $123.3\pm0.1$ & $30.4_{-0.4}^{+0.3}$  &$52.3_{-0.3}^{+0.3}$  & $<0.047        $ & $<13.58        $\\[+0.3ex]             
J111908 & 11:19:06.675 & $+$21:18:29.56 & 0.1383 & 5  &  F606W  & 17.7 &$\cdots$           & $138.0\pm0.2$ & $34.4_{-0.4}^{+0.4}$   &$26.4_{-0.4}^{+0.8}$  & $ 0.074\pm0.005$ & $ 13.83\pm0.02 $\\[+0.3ex]             
J113910 & 11:39:08.330 & $-$13:50:45.64 & 0.2198 & 1  &  F702W  & 22.4 &2.1$^{\dagger}$     & $122.0\pm0.2$ & $44.9_{-8.1}^{+8.9}$   &$85.0_{-8.5}^{+5.0}$  & $ 0.021\pm0.007$ & $ 13.27\pm0.15 $\\[+0.3ex]             
J130112 & 13:01:20.123 & $+$59:01:35.72 & 0.1967 & 1  &  F702W  & 20.9 &1.6$^{\dagger}$     & $135.5\pm0.1$ & $39.7_{-2.2}^{+2.8}$   &$80.7_{-3.2}^{+4.3}$  & $ 0.048\pm0.004$ & $ 13.67\pm0.03 $\\[+0.3ex]             
J131956 & 13:19:55.729 & $+$27:28:12.88 & 0.6719 & 6  &  F702W  & 21.3 & 1.5$^{\dagger}$    & $58.1\pm0.3$  & $22.1_{-5.6}^{+8.5}$   &$15.9_{-8.2}^{+10.1}$ & $<0.054        $ & $<13.64        $\\[+0.3ex]             
J134251 & 13:42:52.235 & $-$00:53:43.10 & 0.2013 & 4  &  F814W  & 19.7 & 2.45$^{\spadesuit}$ & $31.8\pm0.2$  & $44.5_{-0.3}^{+0.1}$  &$71.6_{-0.2}^{+0.3}$  & $<0.037        $ & $<13.47        $\\[+0.3ex]             
J135704 & 13:57:03.290 & $+$19:18:44.41 & 0.4295 & 1  &  F702W  & 21.9 &$\cdots$           & $157.9\pm1.5$ & $8.7_{-1.4}^{+1.6}$    &$85.0_{-1.7}^{+5.0}$  & $<0.033        $ & $<13.42        $\\[+0.3ex]             
J135704 & 13:57:02.914 & $+$19:18:55.51 & 0.4406 & 1  &  F702W  & 21.3 & 1.8$^{\dagger}$    & $141.3\pm0.2$ & $52.9_{-7.6}^{+6.4}$   &$31.3_{-4.1}^{+3.6}$  & $<0.032        $ & $<13.41        $\\[+0.3ex]             
J154743 & 15:47:45.561 & $+$20:51:41.37 & 0.0949 & 1  &  F702W  & 21.4 &1.21$^{\spadesuit}$  & $79.8\pm0.5$  & $54.7_{-2.4}^{+2.0}$  &$80.9_{-2.0}^{+1.8}$  & $<0.038        $ & $<13.48        $\\[+0.3ex]             
J154743 & 15:47:41.642 & $+$20:52:39.39 & 0.1343 & 1  &  F702W  & 19.7 &3.04$^{\spadesuit}$  & $83.8\pm0.1$  & $9.0_{-0.2}^{+0.2}$   &$85.0_{-0.0}^{+5.0}$  & $<0.066        $ & $<13.52        $\\[+0.3ex]             
J170100 & 17:01:03.261 & $+$64:11:59.89 & 0.1900 & 7  &  F814W  & 22.4 &$\cdots$           & $60.6\pm0.2$  & $3.2_{-1.9}^{+1.3}$    &$83.5_{-2.3}^{+1.5}$  & $<0.002        $ & $<13.21        $\\[+0.3ex]             
J170441 & 17:04:37.106 & $+$60:44:20.35 & 0.3380 & 1  &  F702W  & 21.6 &2.0$^{\dagger}$     & $159.4\pm1.1$ & $53.8_{-2.9}^{+3.6}$   &$53.1_{-15.3}^{+7.1}$ & $<0.024        $ & $<13.28        $\\[+0.3ex]             
J213811 & 21:37:48.702 & $-$14:33:16.51 & 0.1857 & 1  &  F702W  & 19.5 &$\cdots$           & $172.8\pm0.4$ & $41.2_{-0.3}^{+0.5}$   &$77.7_{-1.0}^{+1.0}$  & $<0.035        $ & $<13.45        $\\[+0.3ex]             
J213811 & 21:37:45.083 & $-$14:32:06.27 & 0.0752 & 1  &  F702W  & 18.5 &$\cdots$           & $70.9\pm0.7$  & $73.2_{-0.5}^{+1.0}$   &$71.0_{-1.0}^{+0.9}$  & $<0.035        $ & $<13.45        $\\[+0.3ex]             
\cutinhead{Absorbing Galaxies with EW({\OVI})>0.1~\AA}
J012528 & 01:25:28.257 & $-$00:06:08.20 & 0.3787 & 8  &  F702W  & 20.7 & 1.3$^{\dagger}$     & $78.9\pm0.3$  & $57.5_{-3.3}^{+2.4}$   &$68.9_{-1.8}^{+2.0}$   & $0.313\pm0.022$ & $14.57\pm0.06$   \\[+0.3ex]                 
J012528 & 01:25:27.671 & $-$00:05:31.39 & 0.3985 & 8  &  F702W  & 19.7 & 1.8$^{\dagger}$     & $163.0\pm0.1$ & $73.4_{-4.7}^{+4.6}$   &$63.2_{-2.6}^{+1.7}$   & $0.817\pm0.023$ & $15.16\pm0.04$   \\[+0.3ex]                 
J035128 & 03:51:27.892 & $-$14:28:57.88 & 0.3567 & 1  &  F702W  & 20.7 & 0.28$^{\dagger}$    & $72.3\pm0.4$  & $4.9_{-40.2}^{+33.0}$  &$28.5_{-12.5}^{+19.8}$ & $0.396\pm0.013$ & $14.76\pm0.17$   \\[+0.3ex]                 
J040748 & 04:07:49.020 & $-$12:11:20.76 & 0.4942 & 2  &  F702W  & 22.6 & $\cdots$           & $107.6\pm0.4$ & $21.0_{-3.7}^{+5.3}$   &$67.2_{-7.5}^{+7.6}$   & $0.229\pm0.004$ & $14.45\pm0.03$   \\[+0.3ex]                 
J045608 & 04:56:08.913 & $-$21:59:29.00 & 0.4838 & 9  &  F702W  & 20.4 &1.78$^{\dagger}$     & $108.0\pm0.6$ & $85.2_{-3.7}^{+4.4}$   &$42.1_{-3.1}^{+2.7}$   & $0.145\pm0.010$ & $14.16\pm0.15$   \\[+0.3ex]                 
J045608 & 04:56:08.820 & $-$21:59:27.40 & 0.3818 & 1  &  F702W  & 20.7 &1.66$^{\dagger}$     & $103.4\pm0.3$ & $63.8_{-2.7}^{+4.3}$   &$57.1_{-2.4}^{+19.9}$  & $0.219\pm0.013$ & $14.34\pm0.13$   \\[+0.3ex]                 
J091440 & 09:14:41.759 & $+$28:23:51.18 & 0.2443 & 4  &  F814W  & 19.6 &1.24$^{\spadesuit}$   & $105.9\pm0.1$ & $18.2_{-1.0}^{+1.1}$   &$39.0_{-0.2}^{+0.4}$  & $0.333\pm0.028$ & $14.65\pm0.07$   \\[+0.3ex]                 
J094331 & 09:43:30.671 & $+$05:31:18.08 & 0.3530 & 4  &  F814W  & 21.2 &1.17$^{\spadesuit}$   & $96.5\pm0.3$  & $8.2_{-5.0}^{+3.0}$    &$44.4_{-1.2}^{+1.1}$  & $0.220\pm0.024$ & $14.66\pm0.07$   \\[+0.3ex]                 
J094331 & 09:43:32.376 & $+$05:31:52.15 & 0.5480 & 4  &  F814W  & 21.0 &0.96$^{\spadesuit}$   & $150.9\pm0.6$ & $67.2_{-1.0}^{+0.9}$   &$58.8_{-1.1}^{+0.6}$  & $0.275\pm0.050$ & $14.51\pm0.07$   \\[+0.3ex]                 
J095000 & 09:50:00.863 & $+$48:31:02.59 & 0.2119 & 4  &  F814W  & 18.0 &2.74$^{\spadesuit}$   & $93.6\pm0.2$  & $16.6_{-0.1}^{+0.1}$   &$47.7_{-0.1}^{+0.1}$  & $0.211\pm0.019$ & $14.32\pm0.04$   \\[+0.3ex]                 
J100402 & 10:04:02.353 & $+$28:55:12.50 & 0.1380 & 1  &  F702W  & 21.9 &$\cdots$             & $56.7\pm0.2$  & $12.4_{-2.9}^{+2.4}$   &$79.1_{-2.1}^{+2.2}$  & $0.117\pm0.010$ & $14.08\pm0.08$   \\[+0.3ex]                 
J100902 & 10:09:01.579 & $+$07:13:28.00 & 0.2278 & 4  &  F625W  & 20.1 &1.39$^{\dagger}$      & $64.0\pm0.8$  & $89.6_{-1.3}^{+1.3}$   &$66.3_{-0.9}^{+0.6}$  & $0.576\pm0.021$ & $15.14\pm0.10$   \\[+0.3ex]                 
J104116 & 10:41:17.801 & $+$06:10:18.97 & 0.4432 & 10 &  F702W  & 20.9 &2.81$^{\dagger}$      & $56.2\pm0.3$  & $4.3_{-1.0}^{+0.9}$    &$49.8_{-5.2}^{+7.4}$  & $0.368\pm0.023$ & $14.64\pm0.18$   \\[+0.3ex]                 
J113327 & 11:33:28.218 & $+$03:26:59.00 & 0.1545 & 4  &  F814W  & 19.2 &1.29$^{\spadesuit}$   & $55.6\pm0.1$  & $56.1_{-1.3}^{+1.7}$   &$23.5_{-0.2}^{+0.4}$  & $0.252\pm0.026$ & $14.44\pm0.07$   \\[+0.3ex]                 
J113910 & 11:39:11.520 & $-$13:51:08.69 & 0.2044 & 1  &  F702W  & 20.0 &2.30$^{\dagger}$      & $93.2\pm0.3$  & $5.8_{-0.5}^{+0.4}$    &$83.4_{-0.5}^{+0.4}$  & $0.231\pm0.009$ & $14.40\pm0.28$   \\[+0.3ex]                 
J113910 & 11:39:09.801 & $-$13:50:53.08 & 0.3191 & 1  &  F702W  & 20.6 &1.60$^{\dagger}$      & $73.3\pm0.4$  & $39.1_{-1.7}^{+1.9}$   &$83.4_{-1.1}^{+1.4}$  & $0.255\pm0.012$ & $14.41\pm0.09$   \\[+0.3ex]                 
J113910 & 11:39:09.533 & $-$13:51:31.46 & 0.2123 & 1  &  F702W  & 20.0 &2.10$^{\dagger}$      & $174.8\pm0.1$ & $80.4_{-0.5}^{+0.4}$   &$85.0_{-0.6}^{+5.0}$  & $0.137\pm0.009$ & $14.12\pm0.12$   \\[+0.3ex]                 
J121920 & 12:19:23.469 & $+$06:38:19.84 & 0.1241 & 5  &  F702W  & 18.2 &1.20$^{\dagger}$      & $93.4\pm5.3$  & $67.2_{-91.4}^{+39.8}$ &$22.0_{-21.8}^{+18.7}$& $0.424\pm0.020$ & $14.69\pm0.09$   \\[+0.3ex]                 
J123304 & 12:33:04.084 & $-$00:31:40.20 & 0.3185 & 4  &  F814W  & 20.2 &1.38$^{\spadesuit}$   & $30.3\pm0.2$  & $37.1_{-2.3}^{+1.6}$   &$33.8_{-1.2}^{+0.6}$  & $0.439\pm0.021$ & $14.81\pm0.21$   \\[+0.3ex]                 
J124154 & 12:41:53.731 & $+$57:21:00.94 & 0.2053 & 4  &  F814W  & 19.9 &1.42$^{\spadesuit}$   & $21.1\pm0.1$  & $77.6_{-0.4}^{+0.3}$   &$56.4_{-0.5}^{+0.3}$  & $0.519\pm0.018$ & $14.89\pm0.13$   \\[+0.3ex]                 
J124154 & 12:41:52.410 & $+$57:20:43.28 & 0.2178 & 4  &  F814W  & 20.1 &1.53$^{\spadesuit}$   & $94.6\pm0.2$  & $63.0_{-2.1}^{+1.8}$   &$17.4_{-1.6}^{+1.4}$  & $0.366\pm0.017$ & $14.80\pm0.35$   \\[+0.3ex]                 
J124410 & 12:44:11.045 & $+$17:21:05.05 & 0.5504 & 1  &  F702W  & 21.7 &1.34$^{\dagger}$      & $21.2\pm0.3$  & $20.1_{-19.1}^{+16.7}$ &$31.7_{-4.8}^{+16.2}$ & $0.447\pm0.081$ & $14.87\pm0.18$   \\[+0.3ex]                 
J131956 & 13:19:55.773 & $+$27:27:54.84 & 0.6610 & 11 &  F702W  & 21.6 &1.45$^{\dagger}$      & $103.9\pm0.5$ & $86.6_{-1.2}^{+1.5}$   &$65.8_{-1.2}^{+1.2}$  & $0.311\pm0.020$ & $14.55\pm0.06$   \\[+0.3ex]                 
J132222 & 13:22:22.470 & $+$46:45:45.98 & 0.2142 & 4  &  F814W  & 18.6 &2.02$^{\spadesuit}$   & $38.6\pm0.2$  & $13.9_{-0.2}^{+0.2}$   &$57.9_{-0.2}^{+0.1}$  & $0.354\pm0.024$ & $14.62\pm0.12$   \\[+0.3ex]                 
J134251 & 13:42:51.866 & $-$00:53:54.07 & 0.2270 & 4  &  F814W  & 18.2 &1.59$^{\spadesuit}$   & $35.3\pm0.2$  & $13.2_{-0.4}^{+0.5}$   &$0.1_{-0.1}^{+0.6}$   & $0.373\pm0.023$ & $14.58\pm0.11$   \\[+0.3ex]                 
J135704 & 13:57:04.539 & $+$19:19:15.15 & 0.4592 & 1  &  F702W  & 21.4 &1.40$^{\dagger}$      & $45.5\pm0.7$  & $64.2_{-13.8}^{+13.6}$ &$24.7_{-6.5}^{+5.7}$  & $0.419\pm0.026$ & $14.72\pm0.12$   \\[+0.3ex]                 
J155504 & 15:55:05.295 & $+$36:28:48.46 & 0.1893 & 4  &  F814W  & 18.5 &1.43$^{\spadesuit}$   & $33.4\pm0.1$  & $47.0_{-0.8}^{+0.3}$   &$51.8_{-0.7}^{+0.7}$  & $0.385\pm0.033$ & $14.74\pm0.17$   \\[+0.3ex]                 
J213135 & 21:31:35.635 & $-$12:06:58.56 & 0.4300 & 12 &  F702W  & 20.7 &2.06$^{\dagger}$      & $48.4\pm0.2$  & $14.9_{-4.9}^{+6.0}$   &$48.3_{-3.7}^{+3.5}$  & $0.385\pm0.013$ & $14.60\pm0.05$   \\[+0.3ex]                 
J225357 & 22:54:00.417 & $+$16:09:06.82 & 0.3526 & 1  &  F702W  & 20.6 &1.30$^{\dagger}$      & $203.2\pm0.5$ & $88.7_{-4.8}^{+4.6}$   &$36.7_{-4.6}^{+6.9}$  & $0.381\pm0.036$ & $14.70\pm0.15$                              
\enddata 
  \tablenotetext{a}{ 1) \citet{chen01}, 2) \citet{johnson13}, 3) \citet{bychkova06}, 4) \citet{werk12} , 5) \citet{prochaska11}, 6) \citet{churchill12}, 7) \citet{reimers89}, 8) \citet{muzahid15}, 9) \citet{kacprzak10}, 10) \citet{steidel02}, 11) \citet{kacprzak12b}, 12) \citet{gb97}.}
\end{deluxetable*} 


\subsection{Equivalent Width as a Function of Impact Parameter}

In Figure~\ref{fig:D} we show the EW distribution as a function of $D$
for our sample. Note that the dashed horizontal line at 0.1~{\AA} is
our bifurcation between absorbing and non-absorbing galaxies, although
we have shown four detections that reside below this threshold for
completeness only. There is an anti-correlation with EW and $D$, which
is consistent with previous studies that typically show this
anti-correlation between $N$(\OVI) and $D$
\citep{wakker09,prochaska11,tumlinson11,mathes14,johnson15}. A
Kendall-$\tau$ rank correlation test shows that the EW and $D$ are
anti-correlated at the 2.7$\sigma$ level for absorbers with
EW>0.1~{\AA}. If limits are included, then the rank correlation test
shows no correlation (0.9$\sigma$).  The lack of a statistically
significant correlation for the full sample suggests that
non-absorbers exist at all impact parameters and that gaseous halos
have a non-unity covering fraction, even at low $D$. Also shown in
Figure~\ref{fig:D} is the covering fraction as a function of $D$. The
covering fraction is defined as the ratio of the number of absorbers
to the sum of absorbers and non-absorbers in each impact paramater
bin. The covering fraction 1$\sigma$ errors are derived from binomial
statistics \citep{gehrels86}.  The covering fraction starts high at
80\% within $D=50$~kpc and decreasing to 33\% at $D=$200~kpc, which is
consistent with previous {\OVI} studies
\citep[e.g.,][]{wakker09,johnson13}.

\subsection{Equivalent Width as a Function of Galaxy Orientation}


We follow the method of \citet{kacprzak12b} to study the relationship
between {\OVI} absorption and azimuthal angle ($\Phi$). They discuss
that direct binning of the azimuthal angles, an approach taken by some
authors, applies only when the measured uncertainties are smaller than
the bin size and each galaxy can be quantized in single $\Phi$ bin,
which is not always the case. Additional complications occur when the
uncertainties are asymmetric, as they are for GIM2D model output
parameters. We model the measured azimuthal angles and their
uncertainties as asymmetric univariate Gaussian probability
distribution functions \citep[see][]{kato02}, thus creating an
azimuthal angle probability distribution function (PDF) for each
galaxy.  From the continuous azimuthal PDFs, we then compute the mean
PDF combining all galaxies as function of $\Phi$.  The mean PDF
represents the probability of detecting {\OVI} absorption at a given
$\Phi$.  This technique provides higher weight per azimuthal angle bin
for galaxies with well determined $\Phi$. However, even the less
robustly modeled galaxies provide useful information; the method is
similar to stacking low signal-to-noise spectra or images to search
for a coherent signal.

\begin{figure}
\begin{center}
\includegraphics[angle=0,scale=0.9]{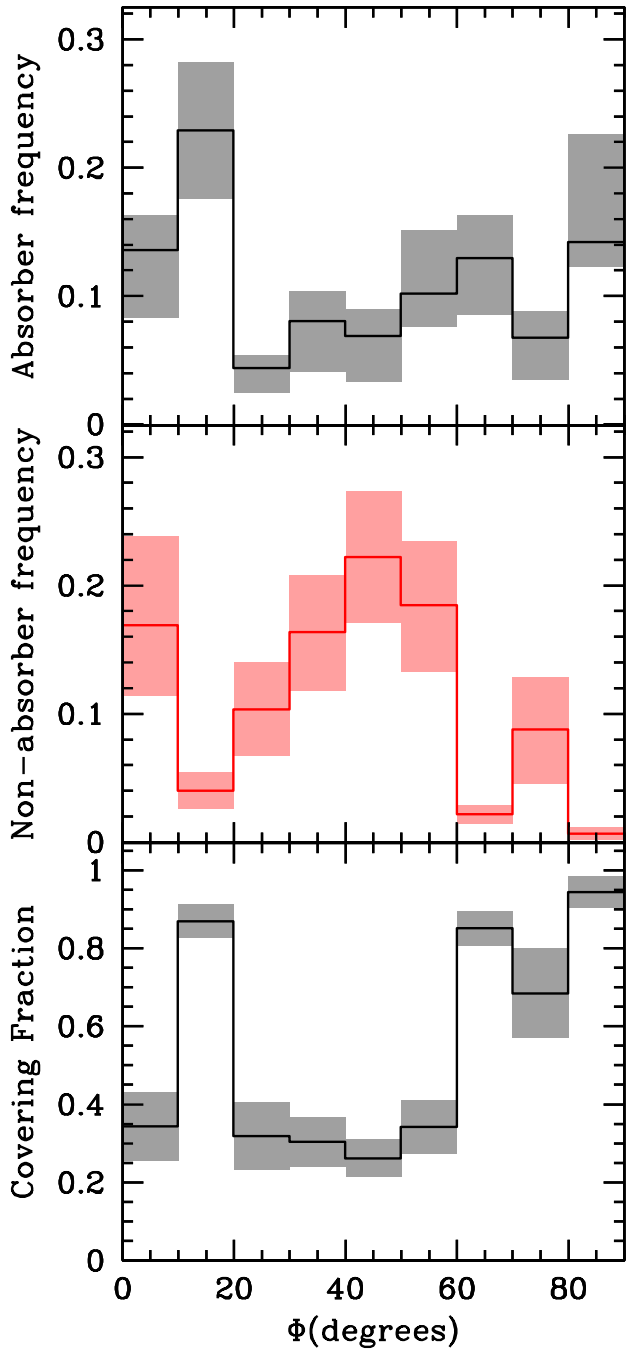}
\caption[angle=0]{(Top) {$\Phi$} distribution for absorbing
  galaxies. The binned PDFs are normalized such that the total area is
  equal to unity: this provides an observed frequency in each
  azimuthal bin. Absorption is detected with increased frequency
  towards the major and minor axes. Shaded regions are 1$\sigma$
  errors produced by bootstrapping the sample. (Middle) {$\Phi$}
  distribution for non-absorbing galaxies. Note the opposite effect,
  where non-absorbers peak where absorbers are near a minimum.
  (Bottom) The {\OVI} gas covering fraction and 1$\sigma$ errors.}
\label{fig:main}
\end{center}
\end{figure}

In Figure~\ref{fig:main}, we present the binned mean azimuthal angle
PDF for the 29 absorbing (EW$>0.1$~\AA) and 24 non-absorbing
(EW$<0.1$~\AA) galaxies. The binned PDFs are normalized such that the
total area is equal to unity: this provides an observed frequency for
each azimuthal bin.  The shaded regions about each bin are the
$1~\sigma$ deviations computed from a bootstrap analysis produced by
resampling the galaxies in the $\Phi$ distribution with replacement
10,000 times.

The top panel of Figure~\ref{fig:main} shows that the PDF for
absorbing galaxies is highest along the projected major axis
($\Phi=0$$^{\circ}$) within 20$^{\circ}$, drops dramatically, and then
slowly rises towards another maximum along projected minor axis
($\Phi=90$$^{\circ}$).  The middle panel of Figure~\ref{fig:main}
shows that the non-absorbers almost have the opposite trend as the
absorbers. The frequency of non-absorbers occur in a narrow window
along the projected major axes within 10$^{\circ}$ and then drops
dramatically.  They have the highest representation at intermediate
$\Phi$ between 20--60$^{\circ}$ where absorbers actually are at a
minimum and have little-to-no representation along the projected minor
axis.

Following the methods of \citet{nielsen15}, we used the chi-squared
statistic on the $10^{\circ}$ binned $\Phi$ distributions to test the
null hypothesis that the distribution of azimuthal angles for
non-absorbers and absorbers were drawn from the same population.  The
null-hypothesis was ruled out at the 3.7$\sigma$ significance level.
To examine sensitivity to binning, we also tested bin sizes of
$2.5^{\circ}$, $5^{\circ}$, and $15^{\circ}$.  The significance levels
were all greater than 3.1$\sigma$.

The data show that the presence or absence of gas is highly driven by
galaxy orientation; absorption is more common along the major and
minor axes while the lack of absorption is common at intermediate
$\Phi$. We discuss how these results are consistent with inflow and
outflow models and how they compare to the {\MgII} azimuthal
dependence seen by \citet{kacprzak12b} in the next section \citep[also
  see][]{bordoloi11,bouche12,bordoloi14}.

We assume that the width of the peaks of the PDF provides constraints
to the geometry of outflowing and inflowing {\OVI} gas.  The peak at
$\Phi=0^{\circ}$ suggests that accreting gas is found within
$\Delta\Phi\simeq\pm20^{\circ}$ of the projected galaxy major axis
plane. One may also interpret that the deficit of absorption and the
surplus of non-absorbers within $\pm$10$^{\circ}$ of the major axis as
physical and could highlight that may be much colder gas may reside in
this area (no such deficiency of absorbers and surplus of
non-absorbers was observed for the cool/dense {\MgII} gas) while more
diffuse gas surrounds the cool dense as indicated by the peak at
20$^{\circ}$. The slow rise of the absorber PDF and the sharp decline
of the non-absorber PDF along the minor axis may suggest that {\OVI}
outflowing gas could occur within a half-opening angle as small as
30$^{\circ}$ (from 60--90$^{\circ}$) or even larger to 50$^{\circ}$
since the covering fraction in outflows likely decreases with
increasing $\Phi$.

In the bottom of Figure~\ref{fig:main}, we further show the {\OVI} gas
covering fraction as a function of azimuthal bin. As defined in the
literature, we define the covering fraction to be $C_f=n_{\rm
  abs}/(n_{\rm abs}+n_{\rm non})$, where $n_{\rm abs}$ is the number
of absorbers and $n_{\rm non}$ is the number of non-absorbers. We
compute the covering fractions and their 1$\sigma$ errors by
bootstrapping the aforementioned ratio for each azimuthal bin 10,000
times.  This first presentation of the {\OVI} covering fraction as a
function of azimuthal angle shows, as previously mentioned, a deficit
of absorption with a covering fraction of 35\% within 10$^{\circ}$ of
the projected major axis. The covering fraction then peaks at 85\% at
20$^{\circ}$ before dropping back to 35\%. The covering fraction of
{\OVI} is again at a maximum above 80\% within 30$^{\circ}$ of the
projected minor axis.  With our current sample, the covering fraction
is highly dependent on $\Phi$ and is the highest along the projected
major and minor axes.  The azimuthal angle averaged covering fraction
of our sample is 52\%.

\begin{figure}
\begin{center}
\includegraphics[angle=0,scale=0.92]{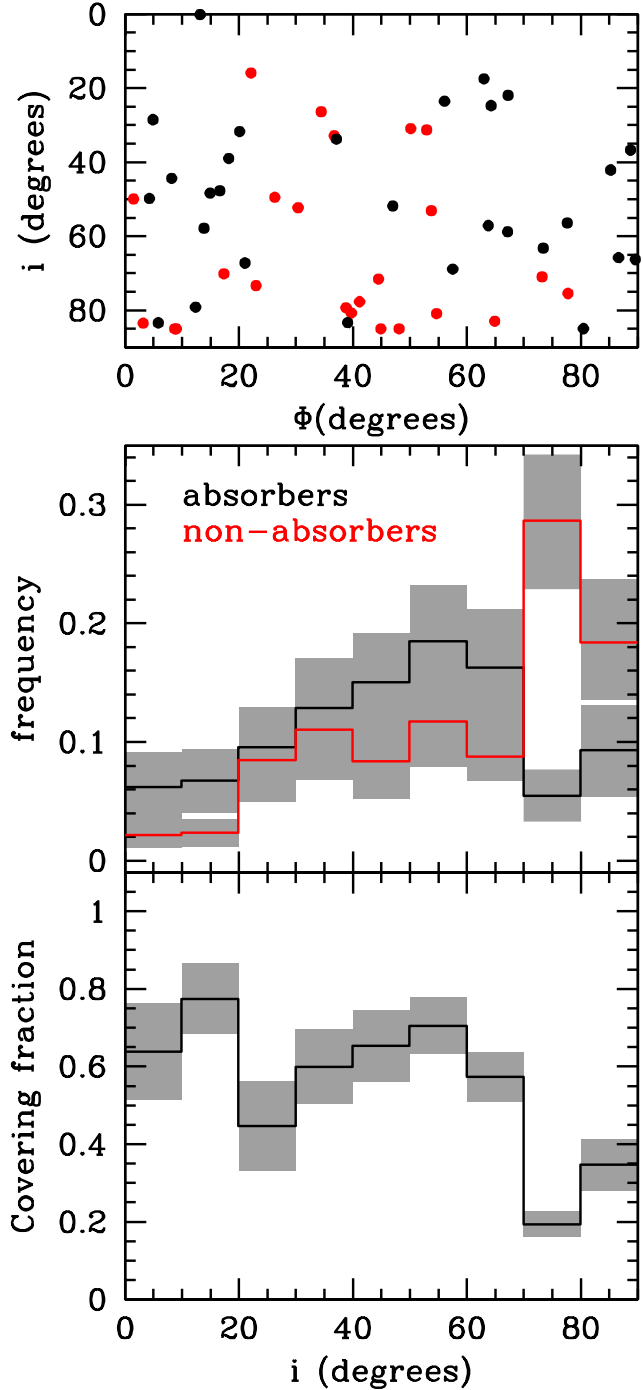}
\caption[angle=0]{ (Top) The distribution of absorbers (black) and
  non-absorbers (red) as a function of position angle ($\Phi$) and
  galaxy inclination ($i)$. Note that the bi-modality in $\Phi$ is
  present for all values of $i$. However, non-absorbers exist over a
  broader $\Phi$ range for edge-on galaxies ($i>70^{\circ}$). ---
  (Middle) $i$ distribution for absorbing and non-absorbing
  galaxies. The binned PDFs are normalized such that the total area is
  equal to unity: this provides an observed frequency in each
  inclination bin. Shaded regions are 1$\sigma$ errors produced by
  bootstrapping the sample. Absorbers and non-absorbers have
  consistent distributions except above ($i>70^{\circ}$). --- (Bottom)
  The {\OVI} gas covering fraction and 1$\sigma$ errors. The drop in
  covering fraction is likely due to a minimized outflow/inflow
  cross-section geometry at high galaxy inclination.}
\label{fig:i}
\end{center}
\end{figure}

We explore how the incidence of absorbers and non-absorbers depend on
galaxy inclination and how the $\Phi$ PDFs depend on inclination.  In
Figure~\ref{fig:i}, we show the distribution of absorbers (black) and
non-absorbers (red) as a function of $\Phi$ and $i$. This Figure shows
that regardless of galaxy inclination, the absorbers have a preference
for existing along the projected major and minor axes. The same is
true for the non-absorbers; they continue to have a preference to
exist at intermediate $\Phi$. We do note that the distribution of
non-absorbers has a larger $\Phi$ range at high galaxy inclination
angles. To further explore this, we show the $i$ PDFs for absorbers
and non-absorbers in Figure~\ref{fig:i}.  The galaxy inclination PDFs
for absorbers and non-absorbers have very similar/consistent
distributions except for the most edge-on systems where a
non-absorbers are more frequent that absorbers. A chi-squared
statistic on the $10^{\circ}$ binned $i$ distributions shows the
null-hypothesis was ruled out at the 3.0$\sigma$ significance
level. However, the chi-squared statistic on the PDFs below
$70^{\circ}$ shows that the null-hypothesis was ruled out at the
1.0$\sigma$ significance level.  Therefore, the largest statistical
difference is observed at high galaxy inclination. This is further
reflected in the covering fractions as a function of galaxy
inclination shown in Figure~\ref{fig:i}. The only significant change
in covering fraction as a function of $i$ is shown as a drastic
decrease at high inclinations.

\begin{figure}
\begin{center}
\includegraphics[angle=0,scale=0.44]{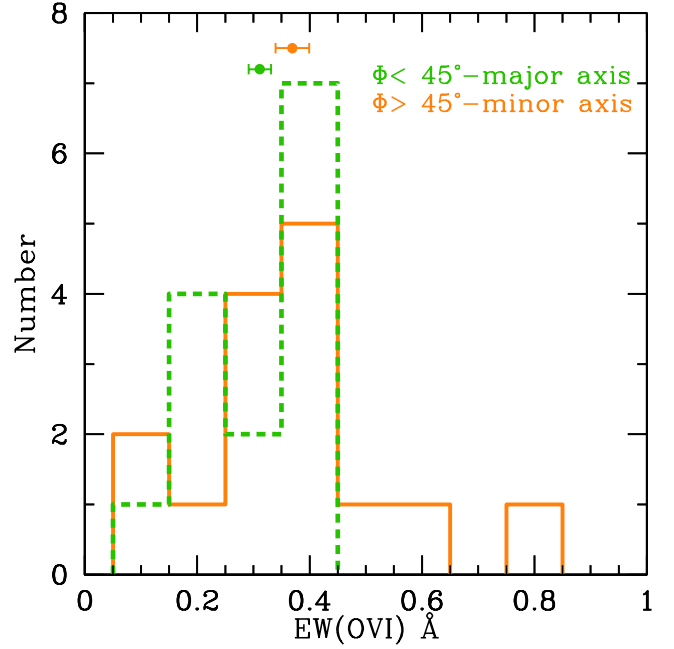}
\caption[angle=0]{The equivalent width distributions for absorption
  detected along the projected major ($\Phi<45^{\circ}$) and along the
  projected minor ($\Phi>45^{\circ}$) axes. The mean value and its
  standard deviation is also shown.  The mean equivalent widths, with
  the standard error in the mean, along the major and minor axes are
  $0.31\pm0.02$~{\AA} and $0.37\pm0.03$~{\AA}, respectively, are
  shown. Note that the highest equivalent width systems are found
  along the projected minor axis.}
\label{fig:diff}
\end{center}
\end{figure}

We further explore how the strength of the absorption is dependent on
the projected location around the galaxy. In Figure~\ref{fig:diff}, we
show the distribution of EW separated into projected major and minor
axis bins bifurcated at 45$^{\circ}$. The EW distributions for the
projected major axis (for 14 galaxies) and minor axis (for 15
galaxies) differ slightly with stronger equivalent width systems found
along the projected minor axis.  The mean equivalent widths, with the
standard error in the mean, along the major and minor axes are
$0.31\pm0.02$~{\AA} (standard deviation of 0.1~{\AA}) and
$0.37\pm0.03$~{\AA} (standard deviation of 0.17~{\AA}),
respectively. Our results suggest that high equivalent width systems
(recall equivalent width is a measure of both kinematic spread and
column density) tend to originate along the projected minor axis from
outflows.  This is consistent with previous results found along the
minor axis for {\MgII} absorption \citep{bordoloi11,kacprzak12a}.  The
same is true for the column densities, with the logarithmic mean
{\OVI} column densities along the major and minor axes being
$14.60\pm0.04$ and $14.73\pm0.06$, respectively.

We also explore how the strength of the absorption is dependent on
galaxy inclination. The mean equivalent widths, with the standard
error in the mean, for low inclination ($i<45^{\circ}$) and high
inclination ($i>45^{\circ}$) galaxies are 0.35$\pm$0.01 (standard
deviation of 0.1~{\AA}) and 0.36$\pm$0.01 (standard deviation of
0.16~{\AA}), respectively. Thus, unlike for high and low $\Phi$, we
find no significant difference in the mean equivalent widths as a
function of galaxy inclination.

\begin{figure}
\begin{center}
\includegraphics[angle=0,scale=0.77]{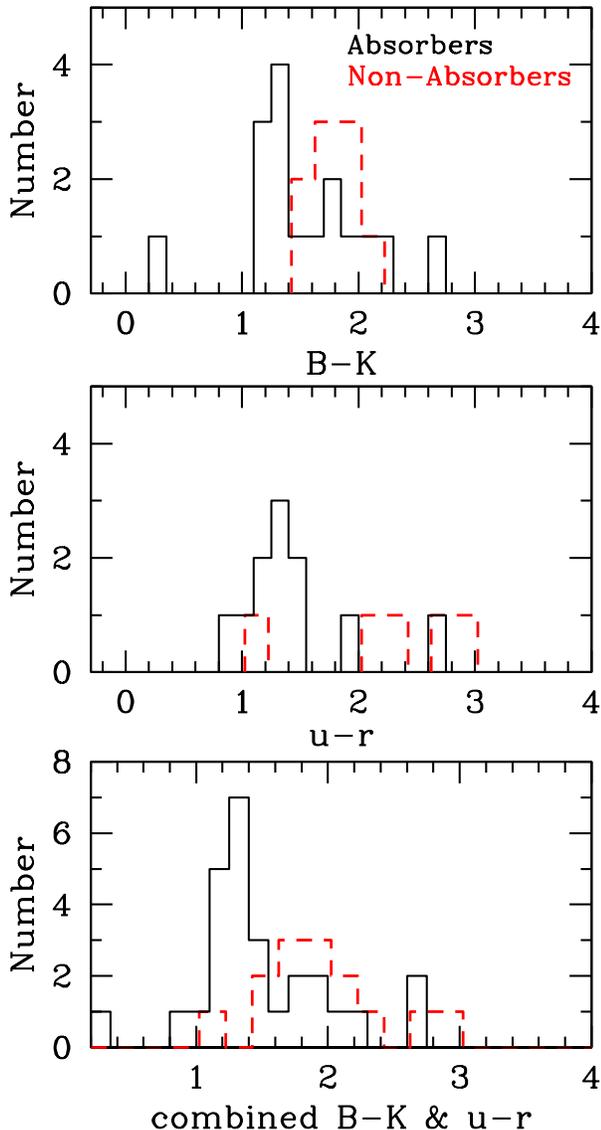}
\caption[angle=0]{(top) $B-K$ color distribution for a subset of our
  sample of absorbing and non-absorbing galaxies. Note non-absorbing
  galaxies tend to be redder than absorbing galaxies. (Middle) $u-r$
  colors for a subset of our sample of absorbing and non-absorbing
  galaxies, again, non-absorbing galaxies are red. (Bottom) Combined
  $B-K$ and $u-r$ colors for 27/29 absorbers and 14/24
  non-absorbers. Note that the bifurcation value between red and blue
  sequences inferred by a color magnitude diagram in $B-K$ and $u-r$
  are both around 2, which makes for roughly an equal
  comparison. Non-absorbing galaxies are redder than absorbing
  galaxies suggesting link between star-formation and halo gas
  cross-section/abundance. }
\label{fig:color}
\end{center}
\end{figure}
\begin{figure}
\begin{center}
\includegraphics[angle=0,scale=0.44]{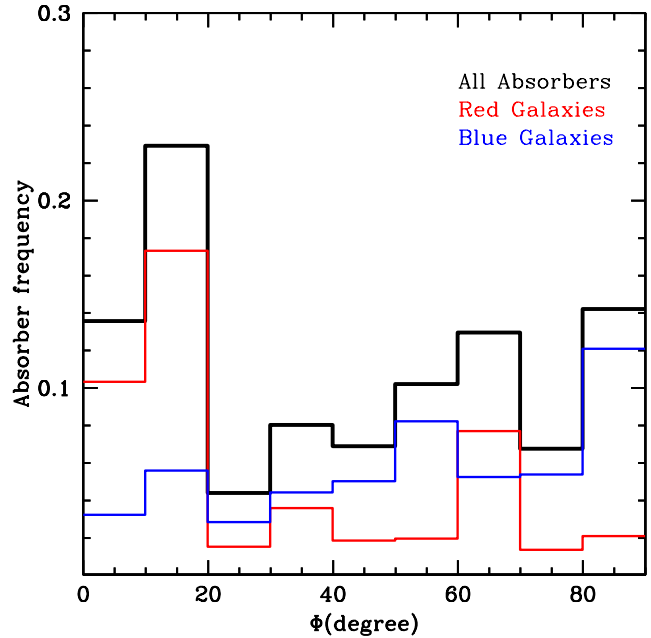}
\caption[angle=0]{{$\Phi$} distribution for absorbing galaxies from
  the top panel of Figure~\ref{fig:main}. We show the relative
  contributions of blue and red galaxies bifurcated at a $B-K$ and
  $u-r$ color of 1.5 (see Figure~\ref{fig:color}). Red galaxies show a
  preference along the projected major axis while blue galaxies prefer
  the project minor axis. This difference in orientation preference
  may explain why blue galaxies are more frequently detected as
  absorbers given the difference in opening angle between inflowing
  and outflowing gas.}
\label{fig:Oricolor}
\end{center}
\end{figure}

\subsection{Galaxy Colors}

Here we assess whether absorbing and non-absorbing galaxies have
different colors.  \citet{tumlinson11} has shown that the majority of
{\OVI} absorbers are produced by star-forming galaxies while
non-absorbers are mostly associated with quiescent galaxies. We do not
have star-formation rates for the majority of our sample, however we
do have $B-K$ or $u-r$ colors for 27/29 absorbers and 14/24
non-absorbers. The $B-K$ colors were either obtained from
\citet{chen01} or from MAGIICAT \citep{nielsen13a} while the $u-r$
colors were obtained from \citet{werk12}. In Figure~\ref{fig:color} we
show the $B-K$ and $u-r$ color distributions for absorbing and
non-absorbing galaxies. Consistent with \citet{tumlinson11}, we also
find the absorbers tend to have bluer colors on average than
non-absorbing galaxies. We combine the $B-K$ and $u-r$ colors in the
bottom panel of Figure~\ref{fig:color}. We validate doing this since
the bifurcation value between red and blue sequences inferred by color
magnitude diagrams in $B-K$ and $u-r$ are both at $\sim2$
\citep[e.g.,][]{gil07,werk12}, thus making for an equal comparison.
The difference in colors between absorbers and non-absorbers is more
accentuated in this combined figure. A KS test shows that absorbers
and non-absorbers differ in their color distribution by 2.5$\sigma$
(CL=0.98793).  This is consistent with the idea that star-forming
galaxies make up a significant fraction of absorbers and a smaller
fraction of non-absorbing galaxies, which are dominated by red
galaxies. This may suggest that blue star-forming galaxies have an
overall higher covering fraction since they may be both accreting and
outflowing gas, while quiescent galaxies have much less CGM activity.

To test our previous statement, we further explore how the different
colors of absorbing galaxies contribute to the distribution of
$\Phi$. We divide 27 absorbers with measured colors into red and blue
galaxies using a color cut at 1.5, which is where the population of
absorbers and non-absorbers split as seen the bottom panel of
Figure~\ref{fig:color}. This color-cut separates blue star-forming
galaxies from redder dusty star-formers, green valley, and red
sequence galaxies. In Figure~\ref{fig:Oricolor}, we show the relative
contribution of absorbers for blue (15 galaxies) and red (12 galaxies)
galaxies. Blue galaxies show a low contribution along the projected
major axis with a gradual increase to peak in frequency along the
minor axis. The red galaxies show the strongest contribution along the
projected major axis with little-to-no contribution along the minor
axis. These results suggest that in fact blue galaxies are dominated
by outflowing gas with wide opening angles while redder galaxies
appear to have {\OVI} along the projected major axis, which has a much
smaller opening angle. This difference in orientation and opening
angle between blue and red galaxies may explain why red galaxies are
less frequently detected due to their low gas sky cross-section while
blue galaxies have a much higher gas cross-section.

\section{Discussion}\label{sec:discussion}

The abundance of {\OVI} contained in the halos of galaxies is
significant \citep{stocke06,tumlinson11,stocke13,peeples14,werk14} and
mostly bound to the galaxy's gravitational potential
\citep{tumlinson11,stocke13,mathes14}. We are just beginning to
understand how this diffuse {\OVI} gas is associated with its host
galaxies and what role it plays in galaxy evolution. We and others
have shown that the covering fraction and equivalent width are
dependent on the distance away from the galaxy
\citep{wakker09,prochaska11,tumlinson11,mathes14,johnson13,johnson15}.
\citet{tumlinson11} has shown that the majority of absorbers are
produced by star-forming galaxies while non-absorbers are mostly
associated with quiescent galaxies. This suggests that either the
quantity of {\OVI} absorbing gas and/or the geometric distribution of this
gas is different between star-forming and quiescent galaxies.

For the first time, we have demonstrated that the distribution of
{\OVI} absorption within 200~kpc of isolated galaxies exhibits an
azimuthal angle bimodality. \citet{mathes14} presented the first hint
of a possible bimodality with only five galaxies detected within one
viral radius with the remaining sample probing larger impact
parameters. Our significant increase in sample size to 29 absorbers
and 24 non-absorbers was key to our findings. Here we have shown that
the highest frequency of absorption is found within 10--20$^{\circ}$
of the projected major axis and within 30$^{\circ}$ of the projected
minor axis. The non-absorbing galaxies occur at the highest
frequencies at intermediate $\Phi$ between 20-60$^{\circ}$. This is
further reflected in the azimuthal covering fraction distribution,
which is typically around 35\% except along the projected major axis
between 10-20$^{\circ}$ and near the projected minor axis between
60-90$^{\circ}$. These results are consistent with the idea that gas
inflows along the major axis and outflows along the minor axis of
galaxies. This bimodality is also consistent with bimodality found for
{\MgII} absorption \citep{bouche12,kacprzak12a}, suggesting the
infalling and outflowing gas is multi-phased. We constrain the half
opening angle of inflowing gas to be around 20$^{\circ}$, which is a
factor of two larger than found for {\MgII} \citep{kacprzak12a}, while
the half opening angle of outflowing gas is at least 30$^{\circ}$,
which is consistent with previous results determined using different
tracers for gas
\citep[e.g.,][]{walter02,kacprzak12a,martin12,bordoloi14}.

Unlike {\MgII}, the covering fraction of {\OVI} within 10$^{\circ}$ of
the projected major axis is very low while the covering fraction of
{\MgII} is above 80\% within 10$^{\circ}$. If this result remains with
a larger sample size, then it may suggest a detection of cool gas
accretion surrounded by an envelope of a diffuse gas phase traced by
{\OVI}. This could have interesting implications for how gas is fed
into galaxies.

\citet{bordoloi11} and \citet{kacprzak12a} reported that projected
minor axis absorption originated from outflows, traced by {\MgII}
absorption, and has higher equivalent widths on average than
absorption found along projected major axis. Here we find the same
result: higher {\OVI} equivalent width systems occur along the
projected minor axis. The mean equivalent widths, with the standard
error in the mean, along the major and minor axes are
$0.31\pm0.02$~{\AA} and $0.37\pm0.03$~{\AA}, respectively. These
results suggest that either the column density, metallicity, and/or
kinetic spread of outflowing gas is higher along the minor axis.
\citet{nielsen15} has shown that {\MgII} absorption profiles with the
largest velocity dispersion are associated with blue, face-on galaxies
probed along the projected minor axis while the column densities are
largest for edge-on galaxies and blue galaxies. These results are
consistent with bi-conical outflows along the minor axis for
star-forming galaxies causing an increase in kinematic spread and
column density. There should also be an increase in metallicity as
well due to enriched supernovae-driven winds but this has yet to be
confirmed observationally. The equivalent width distribution along the
projected major axis still contains moderately high equivalent width
systems. This may suggest that we are not observing purely pristine
accretion onto galaxies and it is likely associated with the recycling
of previously ejected gas
\citep[e.g.,][]{oppenheimer09,oppenheimer10}.

We have shown that the dependence on galaxy inclination is much weaker
than the galaxy position angle and this is consistent with previous
results shown for {\MgII} absorption
\citep{bordoloi11,kacprzak11b,bordoloi14}. We do show that there
appears to be a broader $\Phi$ distribution of non-absorbers for
edge-on galaxies with $i>70^{\circ}$, which translates to a lower
{\OVI} covering fraction for $i>70^{\circ}$. These results are
consistent with the picture that gas inflows along the major axis and
outflows along the minor axis of galaxies. Both the cross-section of
the co-planer flows and the cross-section of conical outflows will be
minimized for edge-on galaxies, thus leading to a higher frequency of
non-absorbers.

Consistent with \citet{tumlinson11}, we find bluer galaxies are
preferentially selected to be absorbers while redder galaxies tend to
be non-absorbers. This suggests that star-formation is a key driver in
the distribution and covering fraction of {\OVI} around galaxies.
However our results shown in Figure~\ref{fig:Oricolor} also suggest
that blue galaxies have {\OVI} detected preferentially along their
minor axes while red galaxies have {\OVI} absorption preferentially
detected along their major axes. This result could arise from our
selected color bifurcation between blue and red galaxies at 1.5 in
$B-K$ and $u-r$, where the ``red'' galaxies could also contain red
star-forming galaxies. However, even if we changed our color cut to 2
(selecting almost only quiescent galaxies), then out of seven red
galaxies six still have $\Phi<17$$^{\circ}$. Thus, the {\OVI}
cross-section of red galaxies could be dominated by small opening
angle accretion of recycled material, while blue galaxies have a
{\OVI} cross-section dominated by large-opening angle outflows. It is
possible then, that the feedback mechanisms of the CGM also drive the
predictions of gas cross-section and not just the presence of
star-formation alone.

\section{Conclusions}\label{sec:conclusion}

We have constructed a sample of 29 {\OVI} absorbing (EW>0.1~{\AA}) and
24 non-absorbing (EW<0.1~{\AA}) galaxies with spectroscopically
confirmed redshifts ranging between 0.08$<$$z$$<$0.67 within
$\sim200$~kpc of bright background quasars. These galaxies are
selected to be isolated such that there are no neighbors within
100~kpc and have velocity separations less than 500~{\kms}. All these
galaxy-absorber pairs were identified as part of our ``Multiphase
Galaxy Halos'' Survey and from the literature. The background quasars
all have medium resolution {\it HST}/COS and STIS ultraviolet spectra
covering the {\OVIdblt} doublet. The quasar fields have high
resolution {\it HST} WFPC-2, ACS, and WFC3 imaging. We used GIM2D to
model the galaxy morphological properties and the azimuthal angle
relative to the galaxy project major ($\Phi=0$$^{\circ}$) and
projected minor ($\Phi=90$$^{\circ}$) axes and the quasar sight-line.
We have analyzed the dependence of absorption on $D$, azimuthal angle,
and $B-K$ and $u-r$ color. Our results are summarized as follows:

\begin{enumerate}

\item We have shown that, for EW$>0.1$~{\AA}, the OVI rest-frame
  equivalent width is anti-correlated with impact parameter, $D$
  (2.7$\sigma$). Non-absorbers (EW$<0.1$~{\AA}) are found at all
  impact parameters, such that the covering fraction is 80\% within
  50~kpc and decreases monotonically to 33\% at 200~kpc.

\item The presence of {\OVI} absorption (EW>0.1~{\AA}) is azimuthally
  dependent and primarily occurs along the projected major axis within
  a half opening angle of 20$^{\circ}$ and along the project minor axis
  within a half opening angle of at least 30$^{\circ}$. These results
  are consistent with what is expected for major axis-fed inflows and
  minor axis-driven outflows as traced by {\OVI} and consistent with
  previous results found for {\MgII} absorption.

\item The frequency of non-detected {\OVI} absorption (EW<0.1~{\AA})
  as a function of azimuthal angle is greatest in the range
  30--60$^{\circ}$. Thus, there is a paucity of detectable OVI
  absorbing gas at intermediate azimuthal angles. This suggests that
  {\OVI} absorbing gas is not mixed throughout the CGM, but remains
  confined primarily within the outflowing winds and near to the
  planar disk region. Non-absorbers also exist within
  $\pm$10$^{\circ}$ of the projected major axis.

\item We show that the covering fraction of {\OVI} is 35\% within
  0-10$^{\circ}$, then peaks at 85\% within 10-20$^{\circ}$, drops
  back to 35\% again between 20-60$^{\circ}$ and peaks up at 80\% from
  60-90$^{\circ}$.  The lack of {\OVI} absorption within 10$^{\circ}$,
  increased presence of OVI absorption at 20--30$^{\circ}$, and then
  sudden decrease beyond 30$^{\circ}$ may suggest that cool gas
  resides in a narrow planar geometry surrounded by warm/hot gas. This
  could indicate accreting gas in which the cool material is narrowly
  confined to the disk plane and is surrounded by an also accreting
  warm/hot envelope, or indicate an extended disk with a pressure
  supported {\OVI} corona, or some combination of both scenarios. The
  covering fraction within 60-90$^{\circ}$ suggests that outflows have
  very high covering fractions.

\item The dependence of absorption on galaxy inclination is much
  weaker than the galaxy position angle and this is consistent with
  previous results shown for {\MgII} absorption. The covering fraction
  is constant at 60\% at all $i$ except it drops by a factor of two at
  $i>70^{\circ}$. This is interpreted as geometric minimization in the
  cross-section of co-planer flows and conical outflows occurring for
  edge-on systems.

\item We determine that the equivalent width distributions for
  projected major axis gas for $\Phi<45$$^{\circ}$ and projected minor
  axis gas for $\Phi>45$$^{\circ}$ have mean equivalent widths, with
  the standard error in the mean, of $0.31\pm0.02$~{\AA} and
  $0.37\pm0.03$~{\AA}, respectively.  Therefore, higher equivalent
  width systems, including the highest EW systems, are found along the
  projected minor axis, which is consistent with an outflow scenario.

\item Consistent with previous results, we show that absorbers tend to
  have bluer colors while non-absorbers tend to have redder colors on
  average. This suggests that star-formation is a key driver in the
  detection rate of {\OVI} absorption.

\item We show the relative dependence of blue (15 galaxies) and red
  (12 galaxies) absorbing galaxies on the distribution of $\Phi$ for a
  color cut of 1.5 in $B-K$ and $u-r$. Blue galaxies show a low
  detection rate along the projected major axis with a gradual
  increase to its peak along the minor axis. The red galaxies show the
  strongest detection rate of absorption along the projected major
  axis with little-to-no detections along the minor axis. These
  results suggest that {\OVI} in the halos of blue galaxies are
  dominated by outflowing gas with wide opening angles while red
  galaxies appear to have gas along the projected major axis, which
  have smaller opening angles. This difference in orientation and
  opening angle between blue and red galaxies may explain why red
  galaxies less frequently produce absorption due to their low gas sky
  cross-section while blue galaxies have a much higher gas
  cross-section.

\end{enumerate}

Our results are consistent with current models of gaseous multi-phase
outflows and accretion/recycling. It is clear that high opening angle
outflows are ubiquitous at all redshifts for star-forming galaxies,
however, it is unclear whether we are still observing small opening
angle cold accretion filaments at redshifts $z<1$ or, the more likely
scenario, we are detecting recycling of previously ejected gas, which
can be seen for most red and a few blue galaxies. Placing constraints
of the metallicity of these absorption systems hold the key in
constraining the exact feedback processes that are occurring in and
around galaxies.


\acknowledgments We would like to thank the referee for his/her
thorough read of the manuscript. We thank, and are grateful to,
Roberto Avila (STScI) for his help and advice with modeling PSFs with
ACS and WFC3. GGK acknowledges the support of the Australian Research
Council through the award of a Future Fellowship (FT140100933). CWC,
JCC, NMN, and SM are supported by NASA through grants HST GO-13398
from the Space Telescope Science Institute, which is operated by the
Association of Universities for Research in Astronomy, Inc., under
NASA contract NAS5-26555.  Based on observations made with the
NASA/ESA Hubble Space Telescope, and obtained from the Hubble Legacy
Archive, which is a collaboration between the Space Telescope Science
Institute (STScI/NASA), the Space Telescope European Coordinating
Facility (ST-ECF/ESA) and the Canadian Astronomy Data Centre
(CADC/NRC/CSA). Some data presented here were obtained at the
W. M. Keck Observatory, which is operated as a scientific partnership
among the California Institute of Technology, the University of
California, and the National Aeronautics and Space Administration. The
Observatory was made possible by the generous financial support of the
W. M. Keck Foundation. The authors wish to recognize and acknowledge
the very significant cultural role and reverence that the summit of
Mauna Kea has always had within the indigenous Hawaiian community.  We
are most fortunate to have the opportunity to conduct observations
from this mountain. Observations were supported by Swinburne Keck
programs 2014A\_W178E and 2014B\_W018E.





{\it Facilities:} \facility{Keck II (ESI)}
\facility{HST (COS, WFPC2, ACS, WFC3)}, \facility{APO (DIS)}.

\end{document}